%% file: 2021gmj.tex
\documentclass[twocolumn]{aastex62}

\usepackage{graphicx}
\usepackage{amsmath}
\usepackage{color}
\usepackage{soul}
\usepackage{CJK}
\usepackage{lineno}

\newcommand{\kms}{{km~s$^{-1}$}}
\newcommand{\rsun}{R$_{\odot}$}

\newcommand{\msun}{$\mathrm{M}_{\odot}$}
\newcommand{\Ni}{\ensuremath{\mathrm{M}_{^{56}\mathrm{Ni}}}}

\graphicspath{{./}{figures/}}

\received{15-Jan-2024}
\accepted{11-May-2024}
\submitjournal{ApJ}

\shorttitle{Low-Luminosity Type II SN~2021gmj}
\shortauthors{Meza-Retamal et al.}

\begin{document}

\title{ Circumstellar interaction signatures in the low luminosity type II SN 2021gmj}

\correspondingauthor{Nicolas E. Meza Retamal}
\email{nemezare@ucdavis.edu}

\input{affiliations.tex}

\input{authors}

\begin{abstract}

We present comprehensive optical observations of SN~2021gmj, a Type II supernova (SN~II) discovered within a day of explosion by the Distance Less Than 40~Mpc (DLT40) survey.
Follow-up observations show that SN~2021gmj is a low-luminosity SN~II (LL~SN~II), with a peak magnitude $M_V = -15.45$ and \ion{Fe}{2} velocity of $\sim 1800 \ \mathrm{km} \ \mathrm{s}^{-1}$ at 50 days past explosion. Using the expanding photosphere method, we derive a distance of $17.8^{+0.6}_{-0.4}$~Mpc. From the tail of the light curve we obtain a radioactive nickel mass of \Ni $= 0.014 \pm 0.001$ \msun. The presence of circumstellar material (CSM) is suggested by the early-time light curve, early spectra, and high-velocity H$\alpha$ in absorption. Analytical shock-cooling models of the light curve cannot reproduce the fast rise,  supporting the idea that the early-time emission is partially powered by the interaction of the SN ejecta and CSM. The inferred low CSM mass of 0.025 \msun \ in our hydrodynamic-modeling light curve analysis is also consistent with our spectroscopy. We observe a broad feature near 4600 \AA, which may be high-ionization lines of C, N, or/and \ion{He}{2}. This feature is reproduced by radiation-hydrodynamic simulations of red supergiants with extended atmospheres. Several LL~SNe~II show similar spectral features, implying that high-density material around the progenitor may be common among them.  

\end{abstract}
\section{Introduction}

\label{sec:intro}
Core collapse supernovae (CCSNe) result from the explosion of massive stars ($>8$~\msun).
When a massive star retains more than $\gtrsim 1$--2 \msun \ of its hydrogen envelope, it explodes as an SN~II (see \citealt{Woosley1994,Sravan2019,Hiramatsu2021a,Gilkis2022}). 
Recent studies have found the photometric and spectral properties of SNe~II to be diverse \citep{Anderson2014,Sanders2015,Galbany2016,Valenti2016,Rubin2016,Pessi2019,deJaeger2019,Hiramatsu2021a}. SNe~II exhibit a large range of peak luminosities, from $-14$ to $-19$~mag in the $V$ band \citep{Anderson2014,Valenti2016}. After the initial rise in luminosity, the onset of hydrogen recombination triggers the plateau phase where the luminosity is roughly constant. During the plateau phase they also show a wide range of photometric decline rates that may correlate with the amount of hydrogen retained by the progenitor pre-explosion \citep[e.g.,][]{Popov1993,Anderson2014,Faran2014,Gutierrez2014,Moriya2016,Hillier2019}. In addition, the light curve properties of SNe~II show heterogeneity at early phases (e.g., rise times, peak brightness, and initial decline rates) that has been attributed  to the interaction of the ejecta with circumstellar material (CSM) \citep{Morozova2017,Morozova2018,Subrayan2023,Hosseinzadeh2023}.

Since the discovery of SN~1997D \citep{deMello1997,Turatto1998}, SNe~II that are particularly faint have been labeled low-luminosity SNe~II (LL~SNe~II). LL~SNe~II have a peak $M_V > -$16 mag \citep{Spiro2014,MullerBravo2020}, along with narrow spectral features (indicating low expansion velocities) \citep{Pastorello2004,Spiro2014}, suggesting explosion energies less than 1~foe\footnote{1 foe = $10^{51}$~erg}. 
 In addition, LL~SNe~II decline more slowly during both the plateau and radioactive-decay light curve phases than other SNe~IIP \citep{Anderson2014,Valenti2016}.

There are two scenarios that explain their observational characteristics. The first results from the explosion of a high-mass star, $\sim25$~\msun, where material remains bound to the core post-collapse, forming a black hole as a remnant. In this scenario the total ejected mass is high, $\sim20$~\msun, but the radioactive yield is severely reduced owing to fallback into the remnant \citep{Zampieri1998}. Secondly, \citet{Chugai2000} proposed a scenario where the progenitor is a low-mass 8--12 \msun \ zero-age main sequence (ZAMS) star. Under this assumption the ejecta mass is lower, $\sim6$--9~\msun, and the radioactive material is meager owing to the low mass of the progenitor's iron core. Progenitor studies of SNe~II using direct imaging have found SNe~II progenitors to be red supergiants (RSGs) with ZAMS masses $\lessapprox20$~\msun \citep{Smartt2009,Smartt2015}. Similar results have been obtained through hydrodynamical models \citep{Dessart2013,Spiro2014,Martinez2019}. The lack of observations of high-mass progenitors favors the aforementioned low-mass progenitor scenario for LL~SNe~II \citep{Spiro2014}. However, this conclusion may be biased by the fact that massive progenitors are intrinsically more rare and that LL SNe~II are more difficult to discover.

The early-time spectra of Type II SNe, within a few days after explosion, may also give us insight into the progenitors. In some instances, early spectra show narrow emission features which quickly disappear within days. The most common interpretation of these ``flash features'' is that they are produced by recombination of CSM after being ionized by the ``flash'' of hard radiation coming from either the shock breakout or the CSM-ejecta interaction \citep[e.g.,][]{Niemela1985, Garnavich1994, Matheson2000, Leonard2000, Quimby2007, GalYam2014, Shivvers2015, Yaron2017,Hosseinzadeh2018,Hiramatsu2021b,Tartaglia2021,Bruch2021}. These features have recently been observed to occur in $\sim30\%$ of SNe~II \citep{Bruch2021}. \citet{Khazov2016} suggests that the presence of flash features is related to SN luminosity, with SNe in their sample only showing flash features for peak magnitudes $M_R < -17.5$~mag; consistent with this, the \cite{Bruch2021} sample found that SNe without flash features are on average fainter. Despite that, recent studies have found LL~SNe~II with early spectral features related to CSM: SN~2016bkv \citep{Hosseinzadeh2018}, which was a LL~SNe~II with flash features, and SN~2018lab \citep{Pearson2023}, which showed a broad emission ledge feature near 4600 \AA that was attributed to CSM interaction. 
Further study on the frequency of spectral features related to CSM in LL~SNe~II is required to understand the role of CSM in the diverse range of progenitors of SNe~II.

In this paper, we present optical observations and analysis of SN~2021gmj, a LL~SNe~II that shows early signatures indicative of CSM.  
Section \ref{sec:obs} outlines the observations and data-reduction procedures. The reddening along the line of sight toward SN~2021gmj is estimated in Section \ref{sec:red}. Section \ref{sec:host} describes the properties of NGC~3310, the host galaxy of SN~2021gmj. In Section \ref{sec:distance}, we estimate a distance to SN~2021gmj with the expanding photophere method (EPM). Section \ref{sec:phot_evo} goes on to describe the photometric evolution, including a multiband and pseudobolometric comparison with other LL~SNe~II. We obtain the radioactive nickel mass in Section \ref{sec:ni}. Following this measurement, in Section \ref{sec:earlylc} we study the early-time light curve with analytic and hydrodynamic models to constrain the progenitor properties. Section \ref{sec:spec_evo} characterizes the spectroscopic observations of SN~2021gmj and describes the overall evolution, together with the early spectral features and possible signatures of CSM interaction. We then derive the low progenitor mass of SN~2021gmj through analysis of nebular spectra in Section~\ref{sec:neb}. In Section \ref{sec:analysis}, we discuss the analysis performed and its implications. Our conclusions are presented in Section \ref{sec:conclusions}.

\section{Observations and Data Reduction}
\label{sec:obs}

\begin{figure}[t]
    \centering
    \includegraphics[width=\columnwidth]{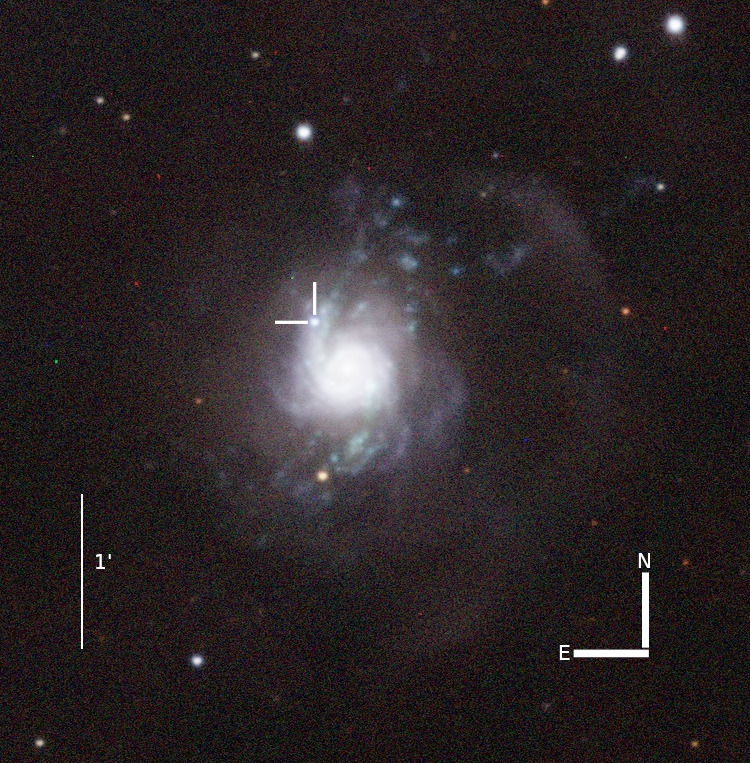}
    \caption{RGB image of SN~2021gmj, marked by white ticks to the northeast of the core of the galaxy NGC~3310, using \textit{gri} images obtained with the Las Cumbres Observatory. The images were taken on 2021 March 20, one day after the estimated explosion of SN~2021gmj.}
    \label{fig:hostImage}
\end{figure}

SN~2021gmj was discovered at RA(J2000) $=10^h38^m47.17^s$, Dec(J2000) $=+53^{\circ}30^\prime31^{\prime \prime}.0$ in the galaxy NGC~3310 (Figure \ref{fig:hostImage}) with the PROMPT-USASK telescope at Sleaford Observatory, Canada on 2021 March 20.32 (UTC dates are used throughout this paper; JD 2,459,293.82) at an apparent $r$ magnitude of 15.98 mag \citep{Valenti2021} as part of the Distance Less Than 40 Mpc Survey \citep[DLT40;][]{Tartaglia2018}. A nondetection $\sim 24$~hr earlier (JD 2,459,292.82), with a limiting magnitude of $r\geq19.1$ mag, constrains the explosion epoch. The SN was independently detected by the Zwicky Transient Facility \citep[ZTF;][]{Bellm2019,Graham2019} on the same night as the reported discovery in the $g$ and $r$ bands at $16.0$ mag and $16.3$ mag, respectively, at JD 24,592,93.74 or 2021 March 20.24 \citep{Zimmerman2021}. Given the well-constrained last nondetection from DLT40 observations, we adopt an explosion date of 2021 March 19.78 throughout this paper. This date corresponds to the midpoint between the last nondetection and the discovery. The explosion epoch and error is then $t_0=$ JD 2,459,293.28 $\pm 0.46$. In Table \ref{tab:sndata} we show basic information and parameters of SN~2021gmj derived in this work. 

\begin{deluxetable}{cc}
\tablecaption{Basic properties of SN~2021gmj used here. \label{tab:sndata}}

\tablecolumns{2}
\tablehead{
\colhead{Property} & \colhead{value} }
\startdata
 \tableline
 RA(J2000) & $10^h38^m47.17^s$ \\
 Dec(J2000)& $+53^{\circ}30^\prime31^{\prime \prime}.0$ \\
 Last nondetection & JD 2,459,292.82 \\
First detection & JD 2,459,293.74 \\
Explosion epoch & JD 2,459,293.3 $\pm 0.5$ \\ 
Host & NGC~3310\\ 
Host Redshift & 0.00331\\  
$E(B-V)_{\rm MW}$ & $0.0192 \pm 0.0005$  mag \\ 
$E(B-V)_{\rm host}$ &  $0.03 \pm 0.01$ mag \\
Distance & $17.8^{+0.6}_{-0.4}$ Mpc \\
$V_{\rm max}$ & $-15.46 \pm 0.08$ mag \\ 
$s_2(V)$  & $0.004 \pm 0.001$ mag $(50\,\mathrm{days})^{-1}$ \\
\tableline
\enddata
\end{deluxetable}

\begin{figure*}[t]
    \centering
    \includegraphics[width=2.0\columnwidth]{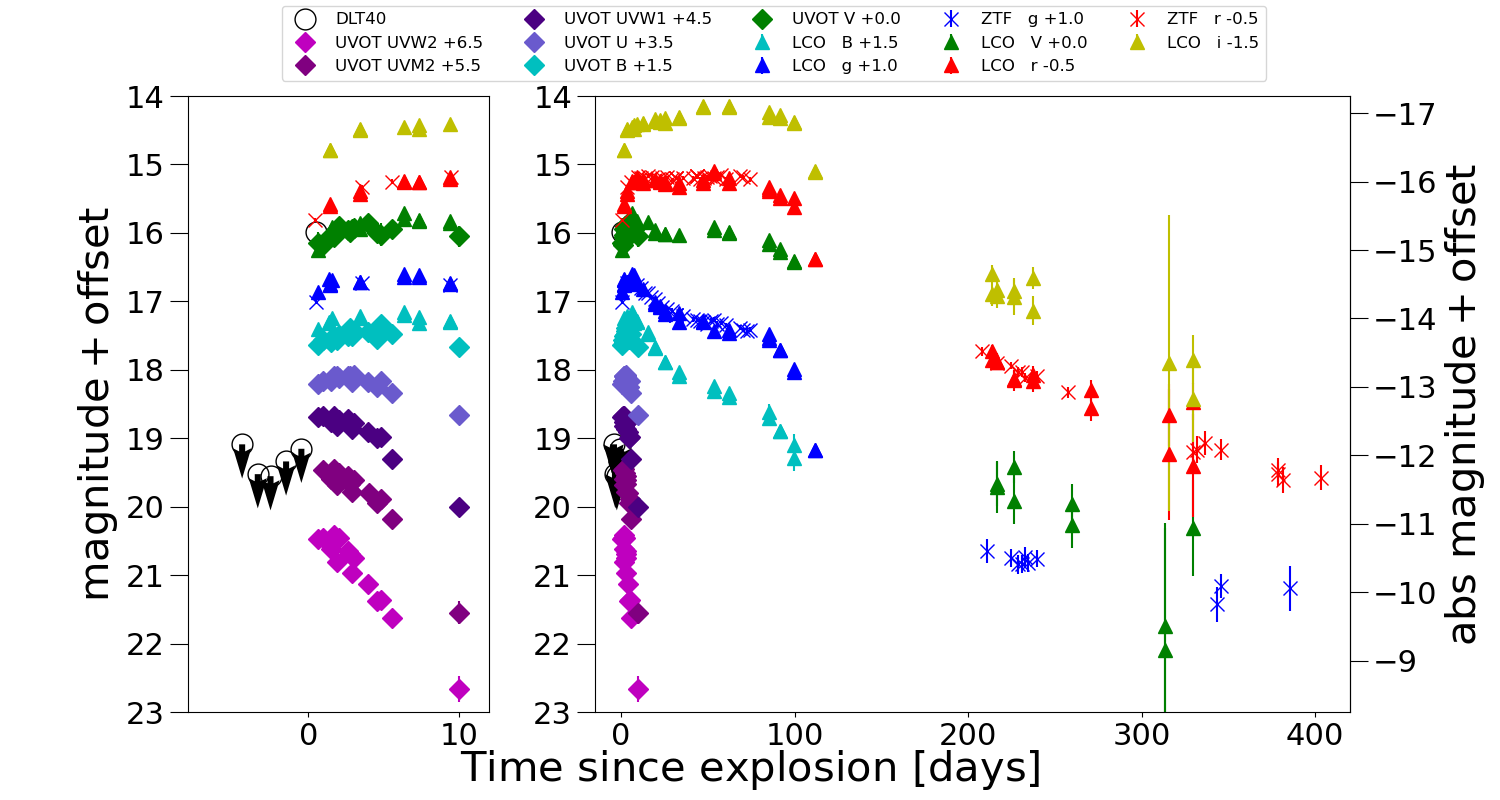}
    \caption{Optical and UV light curves of SN~2021gmj. The left panel contains a zoom-in of the first 10 days of observations, while the right panel shows the full light curves of Las Cumbres $BVgri$ photometry with the addition of ZTF, UVOT and DLT40 detections and nondetections. All the photometry was obtained from difference imaging, which was essential owing to significant host contamination. Pre-explosion nondetections from the DLT40 survey are shown with downward-pointing arrows.}
    \label{fig:lightcurve}
\end{figure*}

Shortly after discovery, high-cadence observations were triggered with the Las Cumbres Observatory \citep{LCOGT} and with the {\it Neil Gehrels Swift Observatory} \citep{Gehrels2004}. Las Cumbres photometric data were reduced using the PyRAF-based pipeline {\sc lcogtsnpipe} \citep{Valenti2016}. Apparent magnitudes were calibrated using the Sloan ($g, r, i$) and APASS ($B, V$) catalogs. Owing to the high background emission from the host at or near the SN location, photometry was obtained after template subtraction using HOTPANTS \citep{hotpants}. The templates used were obtained through the same instruments: Sinistro for the 1~m data \citep{Sinistro} and MuSCAT3 \citep{muscat3} for the single 2~m epoch. {\it Swift} UVOT images were reduced as described by \citet{Brown2009}. The SN coincidence-loss-corrected counts were obtained with a $5''$ aperture radius. The galaxy background coincidence-loss-corrected counts were extracted from the same region using a template image obtained in October 2021. We use updated zero-points that supersede those of \citet{Breeveld2011} to obtain the final calibrated magnitudes, with time-dependent sensitivity corrections updated in 2020\footnote{
\url{https://heasarc.gsfc.nasa.gov/FTP/caldb/data/swift/uvota/}}. Unfiltered ($Open$) DLT40 images were processed with a PyRAF-based pipeline. Background contamination was removed by subtracting a reference image, and the aperture photometry was extracted from the subtracted images. The final photometry is calibrated to the $r$ band using the APASS catalog. Light curves are shown in Figure \ref{fig:lightcurve}. 

Spectroscopic observations of SN~2021gmj started on 2021 March 20, $\sim 1$ day after discovery. The first spectrum was taken with the Liverpool Telescope, which showed SN~2021gmj to be a young SN~II \citep{Perley2021}. 
Another early spectrum was taken with the Ekar Copernico Telescope on 2021 March 21, confirming the classification of SN~2021gmj as an SN~II \citep{Ciroi2021}. Our follow-up spectroscopy started on 2021 March 21 with Binospec \citep{Fabricant2019} on the 6.5~m MMT telescope and the FLOYDS spectrograph \citep{LCOGT} on the 2~m Faulkes Telescope North. We also obtained spectra with the Kast spectrograph on the 3~m Shane telescope at Lick Observatory, as well as with LRIS and DEIMOS on the Keck 10~m telescopes.

The FLOYDS spectra were reduced following standard procedures using the {\sc floyds} pipeline \citep{Valenti2014}. The MMT data were triggered using PyMMT \citep{Shrestha2023} and reduced using the Binospec pipeline \citep{Kansky2019}. Kast spectra were reduced using standard IRAF/Pyraf \citep{pyraf} and Python routines for bias/overscan subtractions and flat-fielding. Finally, LRIS and DEIMOS data were reduced using standard methods with the PypeIt data-reduction pipeline \citep{Prochaska2020}. A summary of spectroscopic observations is given in Table \ref{tab:spectra}.

\begin{deluxetable*}{ccccc}
\tablecaption{Log of optical spectroscopic observations. Phase is relative to our estimated explosion epoch $t_0=$ JD 2,459,293.28 $\pm 0.46$.  \label{tab:spectra}}
\tablecolumns{4}
\tablehead{
\colhead{UTC Date} &
\colhead{MJD} &
\colhead{Phase (days)} &
\colhead{Wavelength range (\AA)} &
\colhead{Telescope/Instrument} 
}
\startdata
\tableline
2021-03-20 & 59293.86 & 1.08 & 4020-7994 & LT/SPRAT \\ 
2021-03-21 & 59294.06 & 1.28 & 3383-8174 & EKAR/AFOSC \\ 
2021-03-21 & 59294.11 & 1.33 & 3827-9198 & MMT/Binospec \\ 
2021-03-22 & 59295.23 & 2.45 & 3200-10,000 & LCO-2m/FLOYDS \\ 
2021-03-23 & 59296.15 & 3.37 & 3827-9198 & MMT/Binospec \\ 
2021-03-26 & 59299.47 & 6.69 & 3200-10,000 & LCO-2m/FLOYDS \\ 
2021-03-28 & 59301.23 & 8.45 & 3200-10,000 & LCO-2m/FLOYDS \\ 
2021-04-01 & 59305.13 & 12.35 & 5688-7210 & MMT/Binospec \\ 
2021-04-02 & 59306.32 & 13.54 & 3200-10,000 & LCO-2m/FLOYDS \\ 
2021-04-05 & 59309.16 & 16.38 & 3620-10,720 & Shane/Kast \\ 
2021-04-08 & 59312.4 & 19.62 & 3200-10,000 & LCO-2m/FLOYDS \\ 
2021-04-14 & 59318.28 & 25.5 & 3622-10,750 & Shane/Kast \\ 
2021-04-14 & 59318.42 & 25.64 & 3200-10,000 & LCO-2m/FLOYDS \\ 
2021-04-17 & 59321.21 & 28.43 & 5688-7210 & MMT/Binospec \\ 
2021-04-18 & 59322.18 & 29.4 & 3632-10,680 & Shane/Kast \\ 
2021-04-22 & 59326.32 & 33.54 & 3200-10,000 & LCO-2m/FLOYDS \\ 
2021-04-28 & 59332.32 & 39.54 & 3200-10,000 & LCO-2m/FLOYDS \\ 
2021-05-09 & 59343.32 & 50.54 & 3642-10,690 & Shane/Kast \\ 
2021-05-15 & 59349.16 & 56.38 & 5207-7703 & MMT/Binospec \\ 
2021-05-18 & 59352.28 & 59.5 & 3628-10,752 & SHANE/KAST \\ 
2021-05-28 & 59362.3 & 69.52 & 3200-10,000 & LCO-2m/FLOYDS \\ 
2021-06-05 & 59370.31 & 77.53 & 3626-10,754 & Shane/Kast \\ 
2021-06-14 & 59296.15 & 86.42 & 3622-10750 & Shane/Kast \\ 
2021-10-11 & 59496.61 & 203.83 & 4401-9126 & DEIMOS/Keck \\ 
2021-11-03 & 59521.5 & 228.72 & 3624-10,720 & Shane/Kast \\ 
2022-01-31 & 59349.16 & 317.26 & 3154-10276 & LRIS/Keck \\ 
2022-06-21 & 59695.34 & 402.56 & 5380-10,311 & LRIS/Keck \\ 
\tableline
\enddata
\end{deluxetable*}

\section{Reddening}
\label{sec:red}
For the Milky Way line-of-sight reddening of SN~2021gmj we use the dust map from \citet{Schlafly2011}; it gives $E(B-V)_{\rm MW}=0.0192 \pm 0.0005$ mag. The medium-resolution MMT spectra taken on 2021 April 1 and 2021 April 17 show \ion{Na}{1}~D lines from the Milky Way and NGC 3310. Host \ion{Na}{1}~D lines, $\lambda$5890 (D2) and $\lambda$5896 (D1), in both medium-resolution spectra were measured, giving equivalent width (EW) values of 0.08~\AA\ and 0.19~\AA\ on 2021 April 1 and 0.07~\AA\ and 0.14~\AA\ on 2021 April 17 for D1 and D2, respectively. Both epoch measurements are consistent within $\sim 25$\% of each other. Using Equation 9 from \citet{Poznanski2012}, we find $E(B-V)_{\rm host}=0.03 \pm 0.01$ mag. Another method for estimating the reddening uses the diffuse interstellar band absorption at 5780~\AA. However, this feature is not seen in any spectrum of SN~2021gmj, consistent with the low reddening derived from \ion{Na}{1}~D lines \citep{Phillips2013}. 

\begin{figure}[h!]
    \includegraphics[width=\columnwidth]{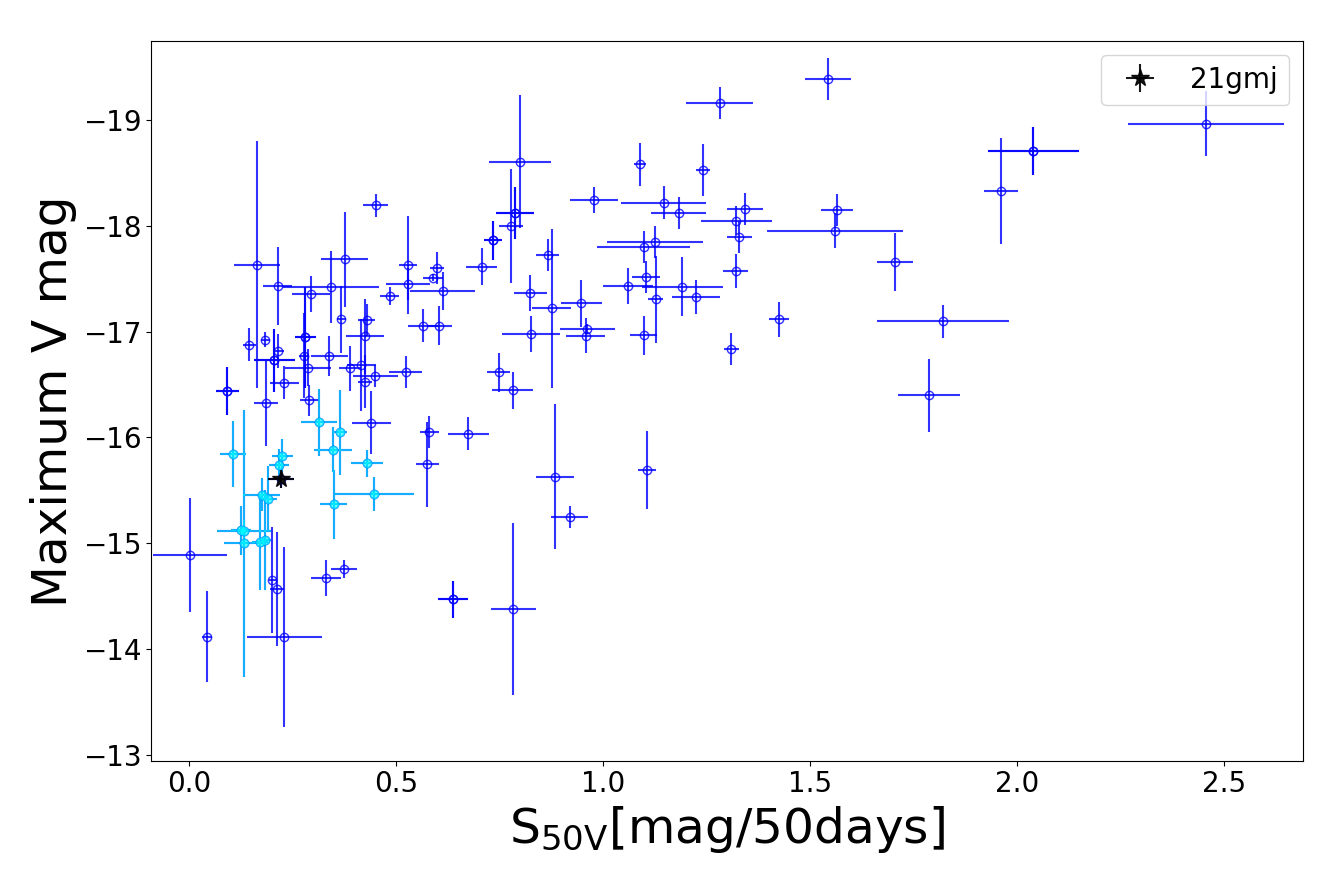}
    \caption{$V$-band light curve slope versus maximum $V$-band magnitude scatter plot for a sample of SN~II \citep[see][]{Valenti2016}. The black star is SN~2021gmj and the cyan data points show the 16 nearest SNe that we use for photometric comparison. Properties and references for the literature sample can be found in Table \ref{tab:photllsne}. \label{fig:s2Vmax}}
\end{figure}

\begin{figure}[h!]
    \centering
    \includegraphics[width=\columnwidth]{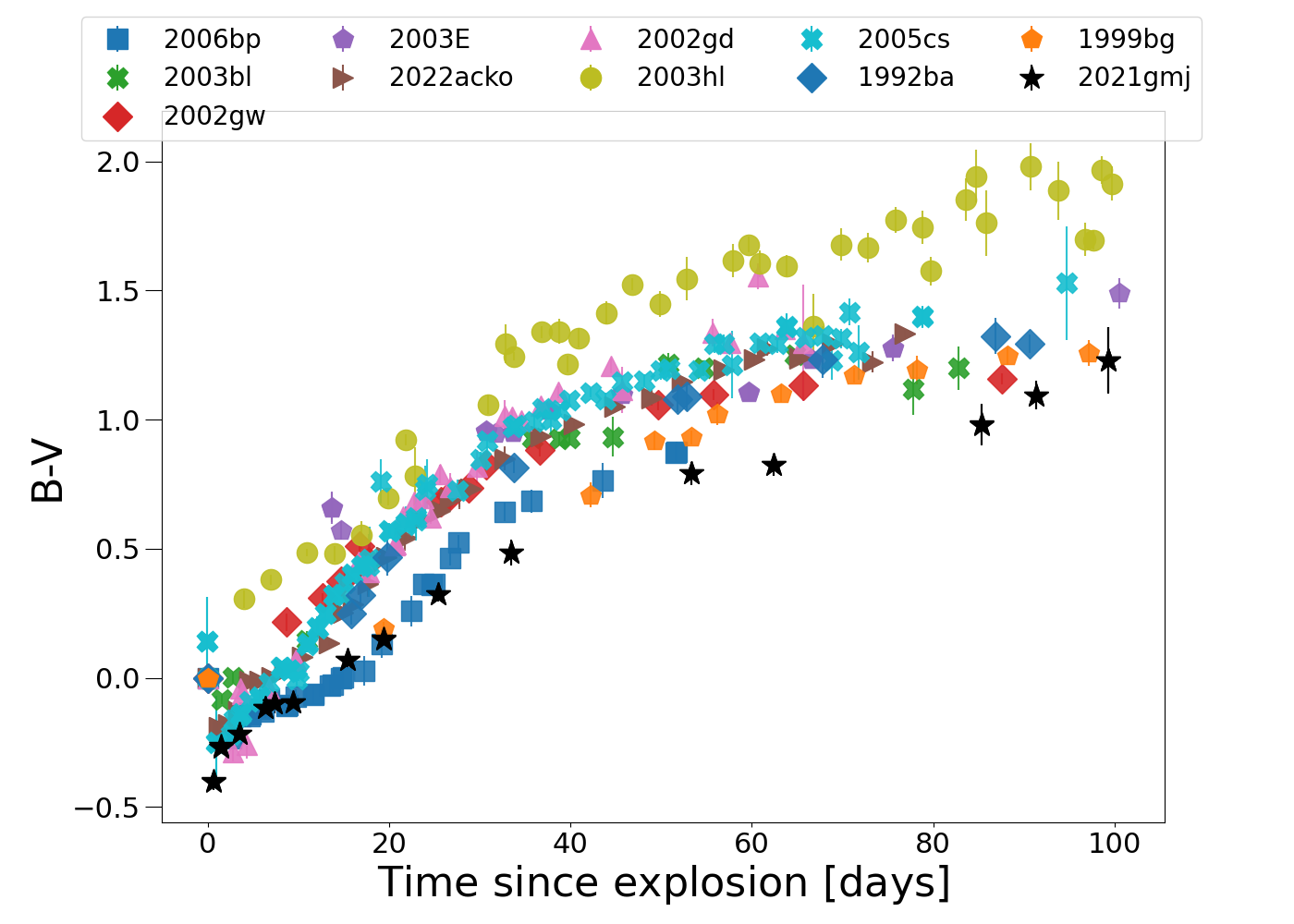}
    \caption{$B-V$ color evolution of SN~2021gmj compared with other SNe~II from the literature. All data have been corrected for host and Milky Way reddening. SN~2021gmj is bluer than the overall sample.}
    \label{fig:color}
\end{figure}

As a sanity check of our estimated reddening, we compared SN~2021gmj colors to those of other SNe~II. The color evolution of SNe~II correlates with the slope of the $V$-band light curve \citep{deJaeger2018}. For this reason we compare the $B-V$ color curve of SN~2021gmj to that of LLSNe~II with similar $V$-band slope. To select the sample we used the Davis supernova database\footnote{\url{http://dark.physics.ucdavis.edu/sndavis/transient}} and retrieved the measured $V$-band slope of the plateau ($s_{50}$) and the $V$-band maximum ($V_{\rm max}$). In this two-dimensional (2D) parameter space of ($s_{50}$,$V_{\rm max}$) we want to select the SNe that are similar to SN~2021gmj. We calculated a ``distance" taking into account the variance in each parameter. This distance, called Mahalanobis distance \citep{Bartkowiak1989,DeMaesschalck2000,Masnan2015}, between two vectors x (in this case x represents the $s_{50}$-$V_{\rm max}$ duple for SN~2021gmj) and y (vector for the comparison SN) is evaluated as
\begin{equation}
    d(x,y) = \sum_{i \in 1,2}{\frac{(x_i-y_i)^2}{\sigma_{x_i}^2}}\, ,
\end{equation}
where $\sigma_{x_i}^2$ is the standard deviation for the corresponding parameter. After measuring the distances between SN~2021gmj and all the SNe we selected the 16 nearest SNe. In Figure \ref{fig:s2Vmax} we show the $s_{50}$ versus $V_{\rm max}$ scatter plot with SN~2021gmj and the comparison sample highlighted. Out of this list of 16 we picked the closest 10 that had the required data to compare the $B-V$ color. Additionally, we included SN~2006bp as it is an object that shows spectral features similar to those of SN~2021gmj (see Section \ref{sec:early_emission}). The resulting color curves can be seen in Figure \ref{fig:color}. SN~2021gmj is the bluest in color evolution, consistent with the low reddening found from the equivalent width of the \ion{Na}{1}~D lines. Because of this, throughout this paper we adopt a total reddening of $E(B-V)_{\rm total}=0.05 \pm 0.01$ mag, as well as an extinction law with $R_V=3.1$ \citep{Cardelli1989}.

\section{Host-Galaxy Properties}
\label{sec:host}
NGC 3310 is a starburst galaxy with a peculiar morphology, showing shell/bow structures and a prominent circumnuclear starburst ring. This galaxy has been extensively studied from X-ray to radio wavelengths. From these studies we know that NGC 3310 has been part of at least a single minor merger, which triggered a burst of star formation that has lasted 10--100 Myr \citep{Elmegreen2002}. This merger enriched the NGC 3310 disk with low-metallicity gas, producing an oxygen abundance of $12+\log(O/H) \approx 8.2-8.4 $ dex in the disk and circumnuclear region \citep{Miralles2014,Hagele2010,Wehner2006,Pastoriza1993}. This translates to sub-solar metallicity in the range between 0.3-0.5 $Z_{\odot}$ \citep[assuming solar abundances from][]{Asplund2009}.  

\begin{figure}[h!]
    \centering
    \includegraphics[width=\columnwidth]{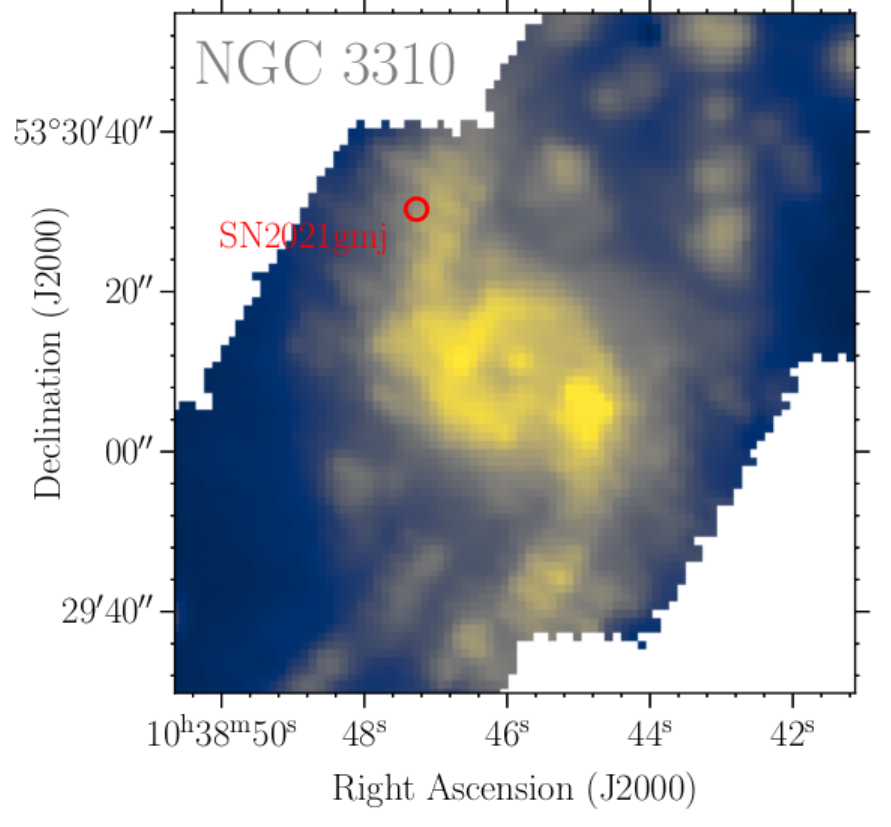}
    \caption{Continuum-corrected H$\alpha$ emission map of NGC 3310. The position of SN~2021gmj is indicated with a red circle. The linear scale corresponds to 86 pc/1 arcsec.}
    \label{fig:host_ha}
\end{figure}

NGC 3310 was observed as part of the PPAK Integral Field Spectroscopy Nearby Galaxies Survey \citep[PINGS;][]{Rosales2010}; details about the observations and reductions are given by \citet{Miralles2014}. The three-pointing mosaic has a spatial sampling of 1 square arcsec per spaxel and the field of view fully covers the SN~2021gmj explosion site. The spatial resolution is dominated by the physical size of the fibers, which translates to a resolution of 2.7 arcsec \citep[More details can be found in][]{Sanchez2016}. A map of the H$\alpha$ emission of NGC 3310 is shown in Figure \ref{fig:host_ha}. To study the properties of the SN~2021gmj explosion site, we follow the method presented in several previous works \citep[e.g.,][]{Galbany2016,Galbany2018}. The key output of the analysis pipeline is a stellar-continuum-subtracted emission-line spectrum of the SN region. The region has a physical size of $\sim 86\times86 \ \mathrm{pc^2}$ on the 1 arcsec$^2$ spaxel (i.e. linear scale of 86 pc/1 arcsec). 

From the spectrum we can obtain key properties of the ionized gas surrounding the SN explosion site. In Table \ref{tab:hostdata} we show the values obtained for the star-formation rate density ($\Sigma$(SFR)), equivalent width of H$\alpha$ (EW($H\alpha$)), reddening as measured by the Balmer decrement, and three different oxygen abundance measurements from the N2, O3N2, and D16 calibrations \citep[][for the N2-O3N2 and D16 calibrations, respectively]{Marino2013,Dopita2016}. Both the SFR intensity and EW($H\alpha$) show the starburst character of the host, with $\Sigma$(SFR) being in the top 20\% of the PISCO sample \citep{Galbany2018} and with the EW($H\alpha$) measurement being the second highest of the whole sample. All of our measurements of the oxygen abundance point to 0.4-0.5 times the solar value \citep{Asplund2009}, which is consistent with previous measurements \citep{Pastoriza1993,Miralles2014}. Our measured value belongs to the lowest 30\% of the PISCO sample. The reddening is higher than our measured value for SN~2021gmj. This may be explained by the larger region probed by a single spaxel at the SN~2021gmj location, which may have on average a higher reddening than the direct line of sight toward SN~2021gmj.

\begin{deluxetable*}{cccc}
\tablecaption{Local properties of the ionized gas near SN~2021gmj as derived from PINGS IFU data, together with the PISCO sample means for comparison; see Section~\ref{sec:host}. \label{tab:hostdata}}
\tablecolumns{4}
\tablehead{
\colhead{Property} & \colhead{value}  & \colhead{unit}
& \colhead{PISCO mean$^*$}}
\startdata
 \tableline
$\Sigma$(SFR)  & 0.065 $\pm$ 0.001  & \msun$\mathrm{yr^{-1} kpc^{-2}}$ & 0.013 \\
E(B-V) & 0.20 $\pm$ 0.02 & mag & --  \\            
EW(H$\alpha$) & 289.2 $\pm$ 0.3 & \AA & 37.4 \\
12+$\log$ O/H (N2) &  8.5 $\pm$ 0.1 & dex & -- \\
12+$\log$ O/H (O3N2) &  8.43 $\pm$ 0.07 & dex & -- \\
12+$\log$ O/H (D16) &  8.40 $\pm$ 0.07 & dex & 8.65 \\
\tableline
(*): If published.
\enddata
\end{deluxetable*}

\section{Distance with the Expanding Photosphere Method}
\label{sec:distance}

The distance to NGC~3310 has available measurements using calibrations of the Tully-Fisher (TF) relationship \citep{deVaucouleurs1981, Bottinelli1984, Giraud1985, Bottinelli1986, Tully1988}. The most recent TF distance is 18.7 Mpc \citep[$\mu=31.36\pm0.40$ mag;][]{Tully1988}{}. While the TF relationship works well for most spiral galaxies, the intrinsic scatter makes the distance to a single galaxy inaccurate \citep{Czerny2018}. Thus, to better constrain the distance to NGC~3310, we use the EPM originally developed by \citet{Kirshner1974}. The EPM is based on a similar method employed for pulsating variable stars from \citet{Baade1926} and assumes that the photosphere is homologously expanding. We can obtain the distance from the relation
\begin{equation}
    t = D\left(\frac{\theta}{v_{\rm phot}}\right ) + t_0\, ,
    \label{epm_d}
\end{equation}
with $D$ the distance, $v_{\rm phot}$ the photospheric velocity, $\theta$ the angular size of the photosphere, and $t_0$ the explosion epoch. Similar to the methodology of previous work \citet{Hamuy2001,Leonard2003,Dessart2005,Jones2009}, the photometry can be combined to write the residuals, 
\begin{equation}
    \epsilon = \sum_{v \in S} \frac{\left \{ m_v-A_v +5log_{10}[\theta \xi (T_c)] - b_v(T_c)\right \}^2}{(\sigma_v^2 + \sigma_{A_v}^2)}\, ,
\label{epm}
\end{equation}
which, after minimizing these residuals, allow us to simultaneously derive the color temperature ($T_c$) and angular size ($\theta$). Here, $\xi$ is the dilution factor, $b_v$ the synthetic magnitude obtained from a blackbody at temperature $T_c$, $A_v$ the extinction, $m_v$ the observed magnitude, and $S$ the filter set. The corresponding uncertainties for the magnitude and extinction are  $\sigma_v$ and $\sigma_{A_v}$, respectively. Here we use the dilution factors from \citet{Jones2009}. 

For this work we use the filter set $BVI$. To obtain $I$ magnitudes from our $ri$ magnitudes we use the Lupton (2005) color transformations\footnote{\url{https://www.sdss3.org/dr8/algorithms/sdssUBVRITransform.php\#Lupton2005}}. The photospheric velocity is estimated by measuring the velocity at maximum absorption of \ion{Fe}{2} $\lambda5169$. Later than approximately 50 days after explosion the relation between $\theta/v$ and $t$ becomes nonlinear \citep{Jones2009}, so we limit ourselves to epochs $< 50$ days from explosion. SN~2021gmj spectroscopy has five epochs in this range where \ion{Fe}{2} $\lambda5169$ velocity was measured by fitting polynomials to the absorption profile of the \ion{Fe}{2} feature. To obtain photometric measurements at each velocity epoch, we do a simple linear interpolation. We implemented a Markov-Chain Monte Carlo (MCMC) code using the Python library \texttt{emcee} \citep{emcee}, that uses the residuals in Equation \eqref{epm} as the log-likelihood together with uniform priors on $\theta$ and $T_c$. To estimate the optimal parameters of Equation \eqref{epm_d} (i.e., the distance and explosion epoch), we use our MCMC code with uniform priors on the explosion epoch $t_0$ and distance $D$. We use a wide prior for $t_0$ which covers 5 days before and 0.5 days after our initial guess: $t_0=2,459,293.28$ JD. Finally, we use a uniform prior between 0 and 100~Mpc in distance. We obtain a distance of $17.8^{+0.6}_{-0.4}$~Mpc and an explosion epoch consistent with our initially assumed value estimated in Section \ref{sec:obs}. This distance is consistent with the TF distances available in the literature. For the remainder of this paper we adopt a distance of $17.8^{+0.6}_{-0.4}$~Mpc to SN2021gmj and therefore to NGC~3310. 

\section{Photometric Evolution}
\label{sec:phot_evo}

The full multiband light curves of SN~2021gmj are shown in Figure \ref{fig:lightcurve}. The light curve evolution resembles that of other LL~SNe~II \citep[Figure \ref{fig:s2Vmax};][]{Gall2015,Valenti2016}. Our early detection and immediate high-cadence follow-up observations capture the multiband light curve rise of 8.4 days in the $V$ band, reaching a maximum brightness of $M_V = -15.45$~mag. Following maximum, the light curve plateaus for about 100 days, with an average slope of 0.004~mag per 50 days in the $V$ band. At the end of the plateau SN~2021gmj drops in brightness by 2 mag in the optical bands. After the fall from the plateau, the light curve settles into the radioactive tail, with a roughly constant slope until our observations stopped around 400 days after explosion.

\begin{figure}[t]
    \centering
    \includegraphics[width=\columnwidth]{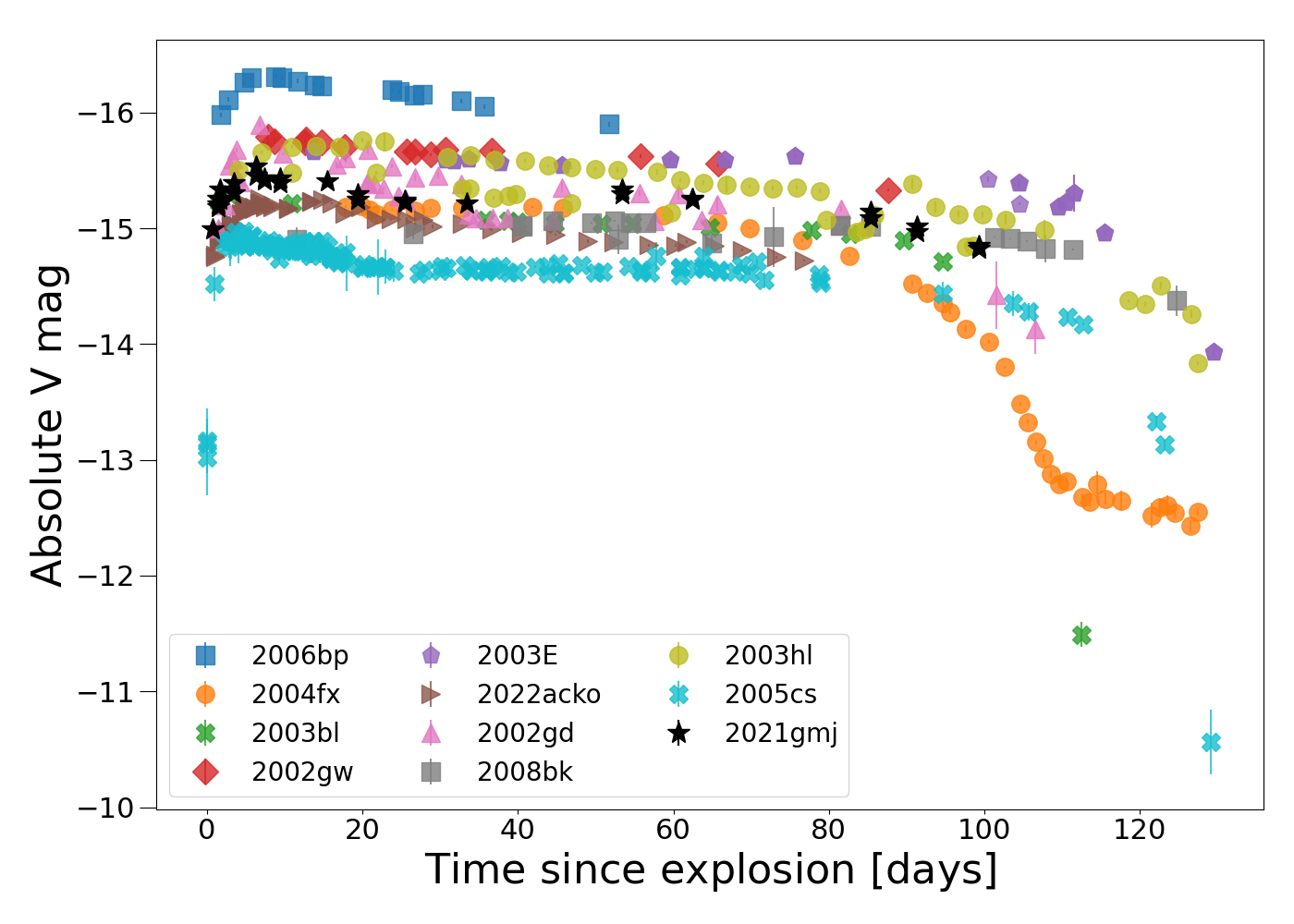}
    \includegraphics[width=\columnwidth]{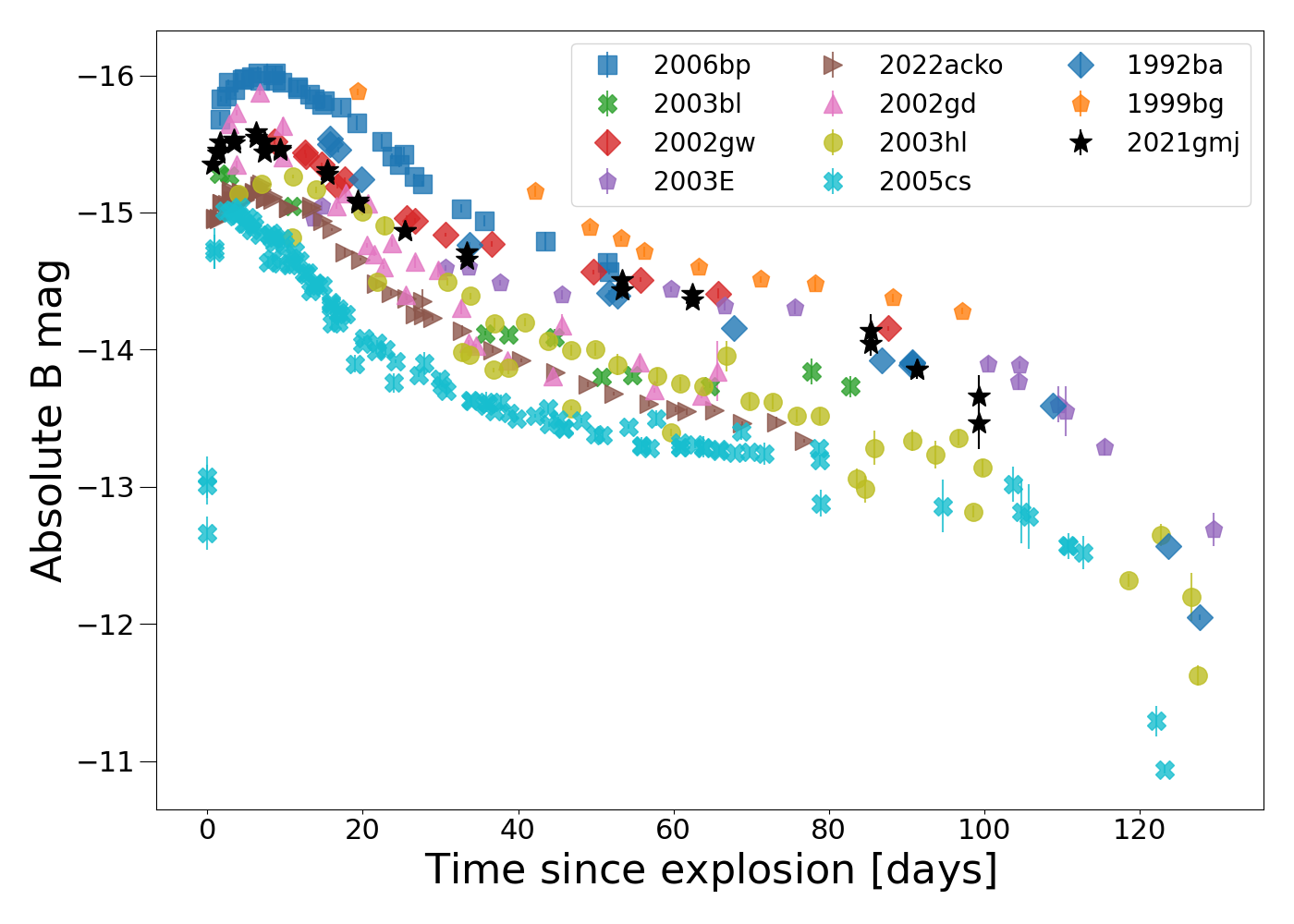}

    \caption{(Top) $V$ light curve of SN~2021gmj compared to a sample of similar type SNe~II. (Bottom) Same for the $B$ band. All light curves show absolute magnitudes, and are dereddened from both Galactic and host extinction. The bluer than average colors of SN~2021gmj can be explained by the brighter $B$ compared to the sample after $\sim50$ days.}
    \label{fig:phot_compare}
\end{figure}

A comparison of the SN~2021gmj $B$ and $V$ light curves with those of other SNe~II taken from our comparison sample is shown in Figure \ref{fig:phot_compare}. 
The relatively bluer color of SN~2021gmj (see Figure \ref{fig:color}) is explained by the $B$ filter brightness which is above the sample average after mid-plateau. The bluer color is also consistent with the weaker iron lines that we describe in Section \ref{sec:spec_evo}. Blue colors have also been associated with CSM interaction in luminous SNe~II \citep{Polshaw2016,Rodriguez2020}. 

\subsection{Pseudobolometric Light Curve}

\begin{figure}[h!]
    \includegraphics[width=\columnwidth]{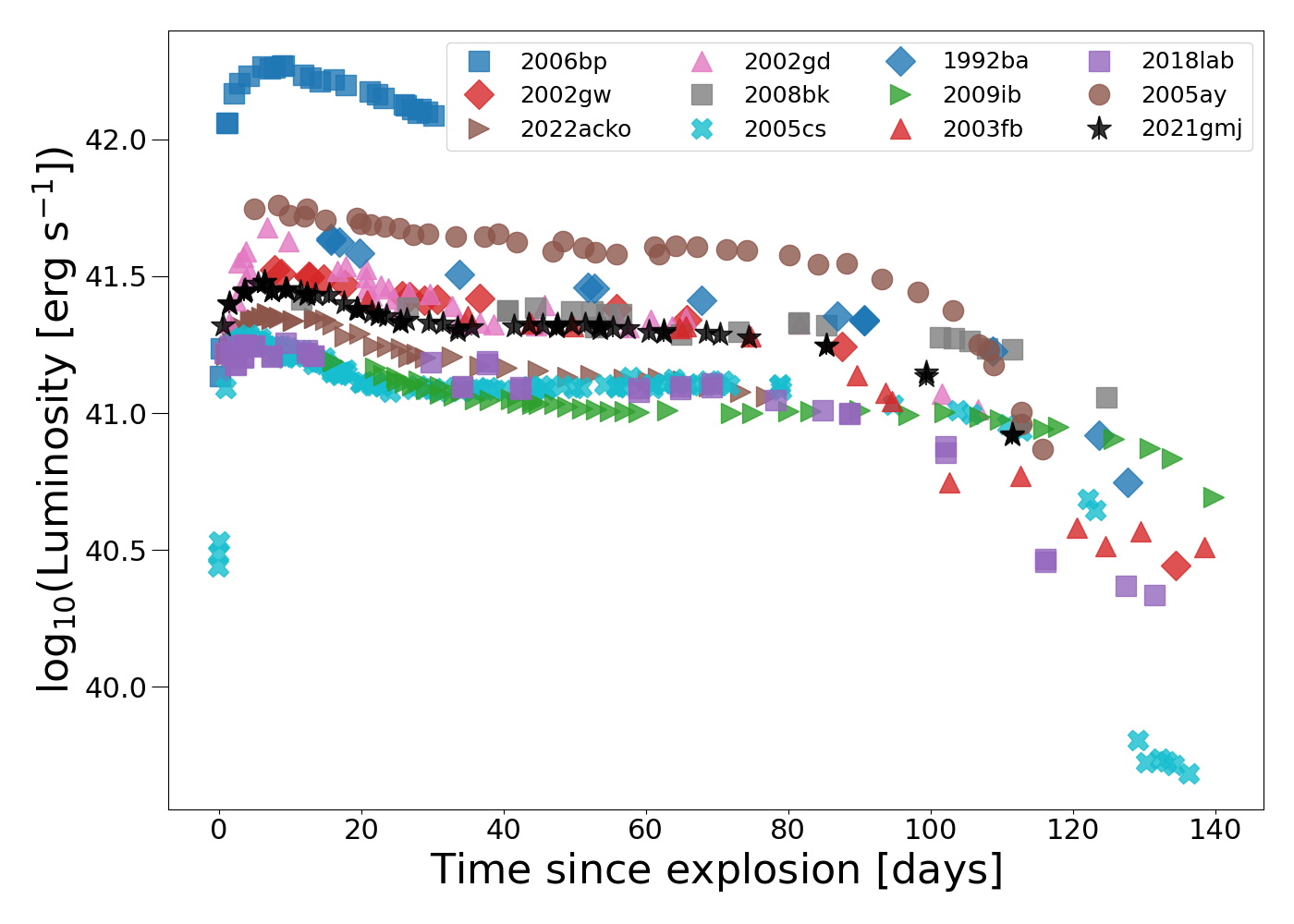}
    \caption{$BVri/BVRI$ pseudobolometric light curves of SN~2021gmj and SNe from the photometric comparison sample, including the more-luminous SN~2006bp. SN~2021gmj fits the average luminosity of the LL~SNe~II sample quite well. The plateau of near 100 days is also expected for this low-luminosity class.\label{fig:Lbol}}
\end{figure}

We construct the $BVri$ pseudobolometric light curve for SN~2021gmj through numerical integration of the spectral energy distribution and compare it with the $BVri$ or $BVRI$ pseudobolometric light curves of our comparison sample; see Figure \ref{fig:Lbol}. 
Despite the blue colors of SN~2021gmj, the bolometric luminosity is similar to the comparison sample in both plateau length and brightness, most closely matching SN~2008bk, SN~2002gw, and SN~2002gd.

\section{Nickel Mass}\label{sec:ni}
During the nebular phase, the light curve is powered by the radioactive decay of \ $^{56}\mathrm{Ni}\Longrightarrow^{56}\mathrm{Co}\Longrightarrow^{56}\mathrm{Fe}$. We calculate the nickel mass of SN~2021gmj by constructing a pseudobolometric light curve from our $Vri$ photometry 200--400 days past explosion. The $B$ observations do not have enough signal-to-noise ratio to detect the SN in the nebular phase. In addition we choose to not include the $g$ band as we need to compare to SN~1987A, which does not have Sloan filter photometry. In addition, we performed several tests to confirm that excluding the $B/g$ bands does not significantly affect our pseudo-bolometric light curve and nickel mass calculation (Bostroem A. et. al. in prep). We then, following \citet{Spiro2014}, compare this integrated luminosity with the pseudobolometric luminosity of SN~1987A in the same bands and epoch. To obtain the Sloan-filters photometry for SN~1987A, we performed synthetic photometry using published optical spectra. With the pseudobolometric luminosity at each epoch, $L_{\rm pbol}$, we fit the following model to obtain the nickel mass (\Ni):

\begin{equation}
    \frac{L_{\rm pbol}(t)}{L^{87A}_{\rm pbol}(t)} = \left(\frac{\Ni}{0.075}\right) \left(\frac{1-\exp(-(T_0/t)^2)}{1-\exp(-(540/t)^2)}\right)\, ,  
\end{equation}

\noindent where $t$ is the time in days after explosion, 0.075 is the nickel mass of SN~1987A in solar mass units, and $T_{0,87A}=540$ days is the gamma-ray trapping timescale for SN~1987A \citep{JerkstrandPhD}. This model assumes that the energy deposition of cobalt decay is diluted by the exponential factor $(1-\exp(-(T_0/t)^2))$ \citep{Clocchiatti1997}. The two parameters of the model, nickel mass ($\Ni$) and trapping timescale ($T_0$), are fitted with an MCMC sampling. We use flat priors from 0 to 1 \msun \ and from 0 to 1000 days for $\Ni$ and $T_0$, respectively. The MCMC chain gives the following means and errors, where the systematic error in \Ni \ is given by the uncertainty in distance modulus: \Ni $= 0.014 \pm 0.001$ \msun \ and $T_0 = 653^{+207}_{-131}$ days are the most probable values. This nickel mass value is consistent with a low-luminosity SN~II \citep[e.g.,][]{Valenti2016}, although it is not as extreme as other SNe with very low-energy explosions \citep[e.g. SN~2005cs $\sim 10^{-3}$ \msun, ][]{Spiro2014}. The ejected nickel is also higher than some SNe from higher-mass progenitors with fallback \citep{Zampieri2003}.

\begin{deluxetable*}{ccccccccc}
\tablecaption{Sample of Low-Luminosity Type II Supernovae. \label{tab:photllsne}
}
\tablecolumns{11}
\tablehead{
\colhead{Name} &
\colhead{Explosion date} &
\colhead{$z$} &
\colhead{$\mu$ (mag)} &
\colhead{$E(B-V)_{\rm MW}$ (mag)} &
\colhead{$E(B-V)_{\rm host}$ (mag)} &
\colhead{$s_2(V)$} &
\colhead{$V_{\rm max}$ (mag)} &
\colhead{References}
}
\startdata
1992ba & 2448889.00 $\pm$ 8.00 & 0.00411 & 31.07 $\pm$ 0.30 & 0.05 & 0.02 & 0.36 $\pm$ 0.02 & -16.05 $\pm$ 0.40 & 1,9,12,16 \\
1999bg & 2451251.50 $\pm$ 14.00 & - & 31.83 $\pm$ 0.32 & 0.02 & 0.00 & 0.32 $\pm$ 0.04 & -16.15 $\pm$ 0.32 & 10 \\
2002gd & 2452552.00 $\pm$ 2.00 & 0.00895 & 32.90 $\pm$ 0.21 & 0.06 & 0.00 & 0.35 $\pm$ 0.05 & -15.89 $\pm$ 0.21 & 9,10,12,16,17 \\
2002gw & 2452560.00 $\pm$ 5.00 & 0.01023 & 33.07 $\pm$ 0.15 & 0.02 & 0.00 & 0.23 $\pm$ 0.03 & -15.83 $\pm$ 0.15 & 9,12,16,17 \\
2003E & 2452635.00 $\pm$ 7.00 & 0.01470 & 34.01 $\pm$ 0.28 & 0.04 & 0.00 & 0.11 $\pm$ 0.03 & -15.85 $\pm$ 0.31 & 9,12 \\
2003bl & 2452700.00 $\pm$ 3.00 & 0.01432 & 34.07 $\pm$ 0.30 & 0.02 & 0.00 & 0.19 $\pm$ 0.02 & -15.42 $\pm$ 0.31 & 9,12 \\
2003fb & 2452777.00 $\pm$ 6.00 & 0.01756 & 34.36 $\pm$ 0.15 & 0.16 & 0.00 & 0.45 $\pm$ 0.10 & -15.48 $\pm$ 0.16 & 9,12 \\
2003hl & 2452869.00 $\pm$ 5.00 & 0.00818 & 32.16 $\pm$ 0.10 & 0.06 & 0.00 & 0.43 $\pm$ 0.04 & -15.76 $\pm$ 0.13 & 9,10,12 \\
2004fx & 2453304.00 $\pm$ 4.00 & 0.00886 & 32.71 $\pm$ 0.15 & 0.09 & 0.00 & 0.18 $\pm$ 0.04 & -15.46 $\pm$ 0.16 & 9 \\
2005ay & 2453453.50 $\pm$ 3.00 & 0.00270 & 31.15 $\pm$ 0.40 & 0.02 & 0.08 & 0.19 $\pm$ 0.03 & -16.32 $\pm$ 0.40 & 2,10 \\
2005cs & 2453549.50 $\pm$ 1.00 & 0.00154 & 29.39 $\pm$ 0.47 & 0.03 & 0.01 & 0.19 $\pm$ 0.01 & -15.03 $\pm$ 0.47 & 2,3,5,6,8,10,16 \\
2006bp & 2453834.50 $\pm$ 0.50 & 0.00351 & 31.42 $\pm$ 0.45 & 0.03 & 0.40 & 0.53 $\pm$ 0.02 & -17.63 $\pm$ 0.46 & 4,5,8,16 \\
2008bk & 2454543.40 $\pm$ 6.00 & 0.00077 & 27.89 $\pm$ 0.23 & 0.02 & 0.00 & 0.13 $\pm$ 0.02 & -15.12 $\pm$ 0.23 & 7,9,16 \\
2009ib & 2455041.80 $\pm$ 3.10 & 0.00435 & 30.32 $\pm$ 0.45 & 0.03 & 0.13 & 0.17 $\pm$ 0.02 & -15.01 $\pm$ 0.45 & 11 \\
2018lab & 2458481.40 $\pm$ 1.00 & 0.00920 & 32.75 $\pm$ 0.40 & 0.07 & 0.15 & 0.13 $\pm$ 0.05 & -15.00 $\pm$ 1.27 & 13,15 \\
2022acko & 2459918.67 $\pm$ 1.00 & 0.00526 & 31.39 $\pm$ 0.33 & 0.03 & 0.03 & 0.35 $\pm$ 0.03 & -15.40 $\pm$ 0.33 & 14,15 \\
\enddata
\tablecomments{Explosion dates and errors are taken from the listed references. References: (1) \citet{Hamuy2001PhD}, (2) \citet{Tsvetkov2006}, (3) \citet{Brown2007}, (4) \citet{Quimby2007}, (5) \citet{Dessart2008}, (6) \citet{Pastorello2009}, (7) \citet{VanDyk2012},   
(8) \citet{Brown2014}, (9) \citet{Anderson2014}, (10) \citet{Faran2014}, 
(11) \citet{Takats2015}, (12) \citet{Galbany2016}, 
(13) \citet{Pearson2023}, (14) \citet{Bostroem2023}, (15) Gaia Photometric Science Alerts, (16) Sternberg Astronomical Institute Supernova Light Curve Catalogue: \url{http://www.sai.msu.su/sn/sncat/}, (17) VSNET: \url{http://www.kusastro.kyoto-u.ac.jp/vsnet/index.html}}
\end{deluxetable*}

\section{Early Light Curve Modeling}

\label{sec:earlylc}

\subsection{Shock-Cooling Models}
The shock-cooling models of \citet{Sapir2017} have been used to characterize the early light curve properties of CC~SNe \citep[e.g.,][]{Hosseinzadeh2018,Andrews2019,Dong2021,Tartaglia2021,Shrestha2023,Hosseinzadeh2023}. These models have been updated in recent work \citep{Morag2023} where an interpolation between planar and spherical phases of expansion is used together with a parameter calibration against hydrodynamic models with a diverse range of progenitor properties. It also includes analytical prescriptions that take into account deviations from a blackbody SED. We follow the prescription of \citet{Hosseinzadeh2018}, and use the MCMC routine implemented in the light curve fitting package \citep{lightcurvefitting}, to constrain the progenitor radius and to identify if shock cooling is the only energy source in the early light curve evolution of SN~2021gmj. 

\begin{figure*}[t]
    \centering
    \includegraphics[width=0.9\textwidth]{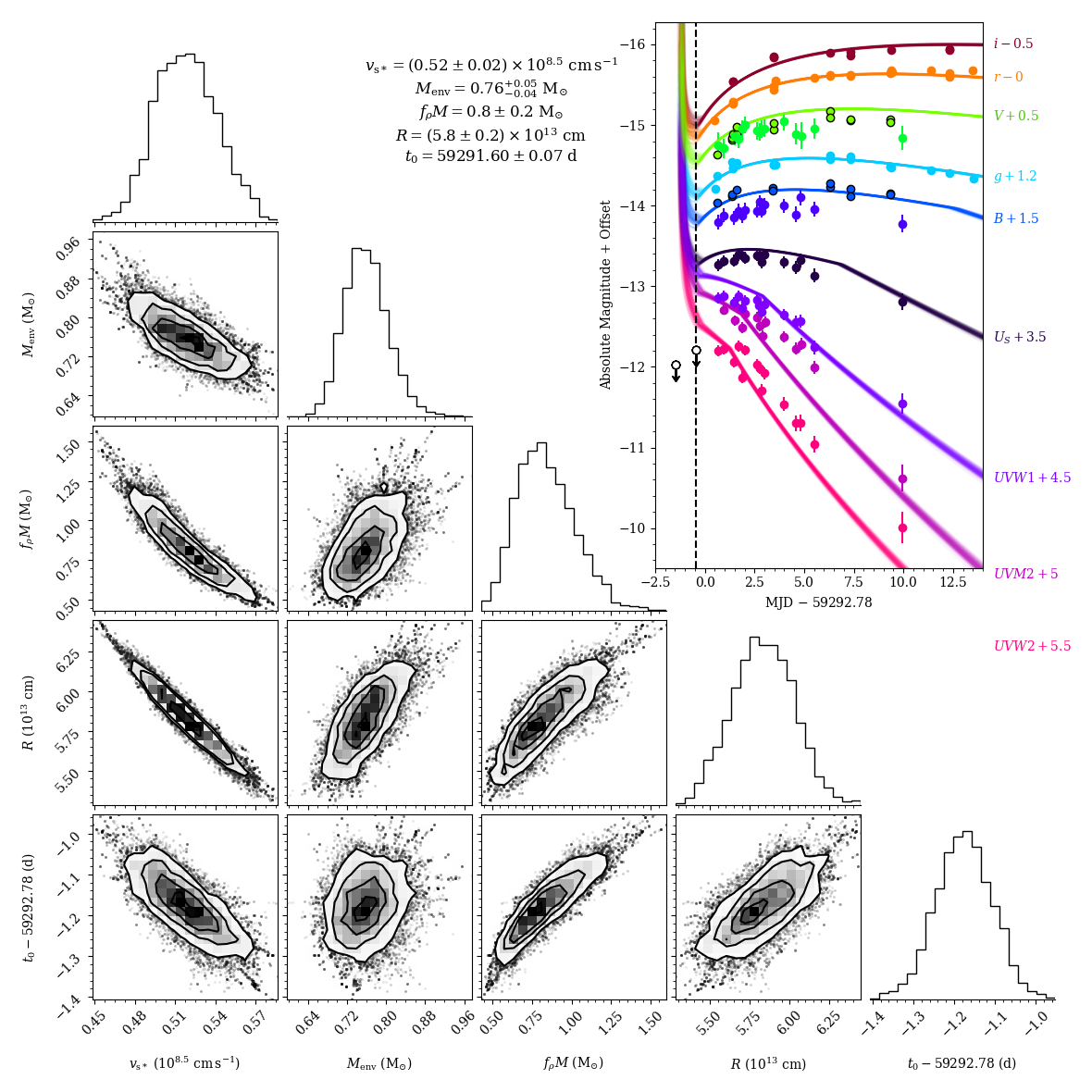}
    \caption{Posterior probability distributions for the progenitor radius $R$, shock velocity scale ($v_{s*}$),  envelope mass ($M_{\rm env}$), scaled ejecta mass ($f_{\rho}M$), and time of explosion ($t_0$). The top-right panel shows a sample of model light curves using randomly selected parameters drawn from the posterior, together with the observed magnitudes. The last two nondetections are also shown with white circles, and a dashed line also shows the last nondetection for clarity. The resulting median and $1\sigma$ for the best-fit parameters are listed at the top of the figure. The model does not fit well the UV bands  and it violates our nondetections, indicating that the model is not able to fit the fast rise in the light curves. We interpret this as evidence of CSM interaction.}
    \label{fig:shockcooling}
\end{figure*}

We use our multiband ultraviolet (UV) plus optical light curve up to 15 days post-explosion to fit for the model. The model assumes a uniform density core and an $n=3/2$ polytrope envelope suitable for an RSG. An MCMC routine is used to fit for the model parameters: the progenitor radius ($R$), a shock velocity scale ($v_{s*}$), the envelope mass ($M_{\rm env}$), the product of the total ejecta mass ($M$) and a constant of order unity ($f_{\rho}$) [hereafter ``scaled ejecta mass'' ($f_{\rho}M$)], and the time of explosion ($t_0$). 
Flat priors are assumed for each parameter (see \citealt{Hosseinzadeh2018} for a complete description of the codes and assumptions used), and in the case of the explosion epoch we take a flat prior $-3.0$ days to 0 days from discovery. The resulting posterior distributions and fits are shown in Figure \ref{fig:shockcooling}.

The best-fit radius of $5.8^{+0.2}_{-0.2} \times 10^{13} \ \mathrm{cm} \approx 834 \ R_{\odot}$ is consistent with typical RSG radii \citep{Levesque2005}. Other parameters of the model ($M_{\rm env}$, $f_{\rho}M$, and $v_{s*}$) cannot be directly compared with the progenitor and explosion properties as these parameters depend on the density structure of the progenitor, which is controlled by the parameter $f_\rho$ and not constrained by the model. The model fits neither the UVOT bands nor the optical bands. In particular the UV bands of the model underestimate the emission. Although the discrepancy is not high considering the light curve dispersion, this may be an indication that the early emission requires an extra source of energy \citep{Hosseinzadeh2018,Dong2021,Pearson2023,Shrestha2023}. Other signs of the struggle of the model to do a proper fit are seen in Figure \ref{fig:shockcooling}, where we show how the model is inconsistent with our nondetections by pushing the explosion epoch to lower values. This can be interpreted as the inability of the model to have a fast rise; it has also been interpreted as evidence of CSM interaction in previous work. In the next section we will find more evidence for the presence of material surrounding the progenitor of SN~2021gmj.

\subsection{Hydrodynamic Modeling of the Light Curves with SNEC}
\label{sec:SNEC}
To further constrain progenitor properties, and to understand if any contribution of the early-time light curve is coming from CSM interaction, we use the open source 1D hydrodynamic code, SuperNova Explosion Code (\texttt{SNEC}) \citep{Morozova2015}. 
\texttt{SNEC} takes as input a progenitor model (density, temperature, velocity, etc.), an explosion energy $E$, a nickel mass, and a nickel mass outer boundary. 
Following \citet{Morozova2017}, we fix the nickel mass boundary up to 5 \msun\ as this weakly affects the light curve. 
We also fix the nickel mass to be \Ni $= 0.014 \pm 0.001 {\rm (stat)} \pm 0.001 {\rm (sys)}$ \msun \, as obtained in Section \ref{sec:ni}. 

To convert bolometric properties of the model to broad-band photometry, we assume a black-body spectral energy distribution. As the actual spectra of SNe show significant line-blanketing from metal lines, a black-body fit is not appropriate for short wavelengths. Owing to this we exclude filters bluer than $g$. We also do not include the $V$ band as it overlaps significantly with the $g$ and $r$ wavelength ranges. 

As done by \citet{Morozova2018,Dong2021}, we use a two-step approach to fit our $gri$ optical light curves. We use the solar metallicity progenitor models of \citet{Sukhbold2016} which were calculated with the stellar evolution code \texttt{KEPLER} \citep{Kepler}. We first constrain the explosion energy $E$ and progenitor mass $M$ by fitting the plateau between 27 and 112 days after explosion. These dates correspond to the end of the initial slope \citep[
][]{Anderson2014,Valenti2016} and the end of our data during the fall from the plateau. We consider only RSG progenitors, using models with masses in the range 9--15 \msun\ in steps of 0.5 \msun\ and 17, 19, and 21 \msun. Our grid of explosion energies ranges from 0.01 to 0.5 foe, with 20 points equally spaced, and we add 4 equally spaced points from 0.6 to 1.4 foe. We define the best-fit $E$ and $M$ as the parameters which minimize the $\chi^2$ value over all three bands considered at all epochs. The model with the lowest $\chi^2$ has $M=10$ \msun\ and $E=0.294$ foe, although we remark that the models can fit with a similar probability using both a lower mass and smaller explosion energy. 

\begin{figure}[h!]
    
    \includegraphics[width=\columnwidth]{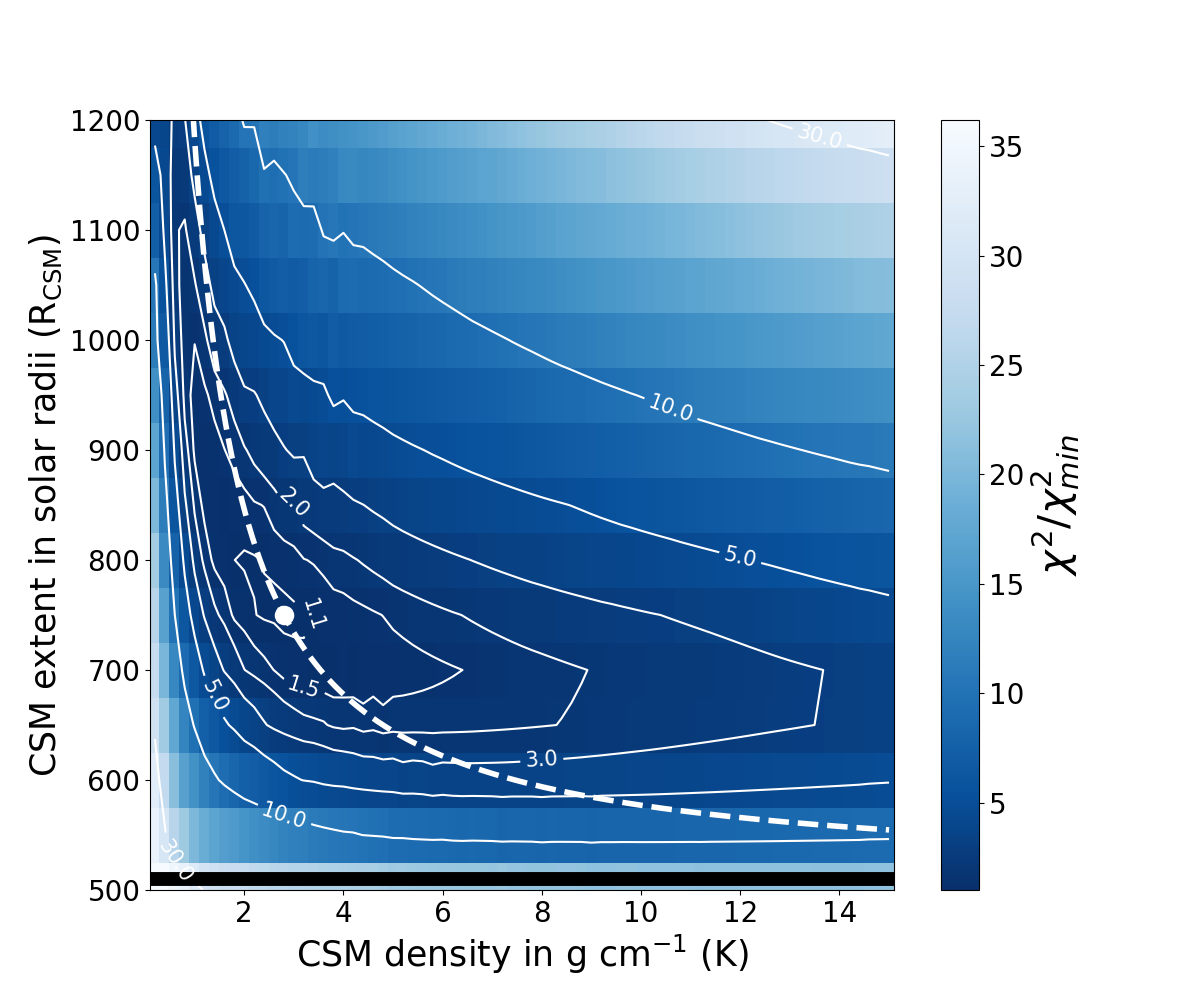}
    \caption{$\chi^2$ surface plot for the CSM models as discussed in Section \ref{sec:SNEC}. The best-fit parameters are marked with a white dot, which corresponds to a density of $K=2.4\times 10^{17} \ \mathrm{g} \ \mathrm{cm}^{-1}$ and a CSM radius extent of $R=750$ \rsun. Contour lines are also shown and the black thick line denotes the progenitor extent without CSM (510 \rsun, KEPLER model). There is a clear degeneracy between both parameters, which was also observed in other works \citep[e.g.,][]{Morozova2017}. The degeneracy follows closely the curve where the CSM mass is constant (white-dashed line in the plot), at a value of $M_{\rm CSM} = 0.025$ \msun\ at the best fit.}
    \label{fig:CSMchi2}
\end{figure}

After fixing $E$ and $M$, we explore the influence of CSM by superimposing a steady wind on the original density profile. This wind extends from the progenitor radius $R_*$ up to a radius $R_{\rm CSM}$, 
\begin{equation}
    \rho(r) = K/r^2 \ ; \ R_* \leq r < R_{\rm CSM}\, ,
\end{equation}
where $K$ is the line density that can also be written as $K = \dot{M}/v_{\infty}$, where $\dot{M}$ is the mass-loss rate and $v_{\infty}$ is the velocity of the CSM. We run models with an outer CSM radius between 500 and 1200 \rsun \ in steps of 50 \rsun \ and density from $0.2 \times 10^{13} \ \mathrm{g} \ \mathrm{cm}^{-1}$ to $15 \times 10^{13} \ \mathrm{g} \ \mathrm{cm}^{-1}$ in steps of 0.2 $\mathrm{g} \ \mathrm{cm}^{-1}$. We again calculate the reduced $\chi^2$  value for each model to obtain the best-fit parameters, but this time we only compare the light curve between the explosion epoch and the end of the light curve slope break ($s_1$ phase) at 26.72 days. In Figure \ref{fig:CSMchi2} we show the surface plot of the $\chi^2$ values in our $R$--$K$ parameter space. Our best-fit model has $K=2.4\times10^{13} \ \mathrm{g} \ \mathrm{cm}^{-1}$ and $R=750$ \rsun, and there is a clear degeneracy between both parameters. This has been observed in previous work \citep[e.g.,][]{Morozova2018} and indicates that the model more strongly constrains the total CSM mass, which is a product of $K$ and $R_{\rm CSM}$ in this simple model:
\begin{equation}
    M_{\rm CSM} = \int_{R_*}^{R_{\rm CSM}}{4\pi\rho r^2 dr} = 4\pi K (R_{\rm CSM}-R_*)\, . 
\end{equation}

\begin{figure}[h!]
    \includegraphics[width=\columnwidth]{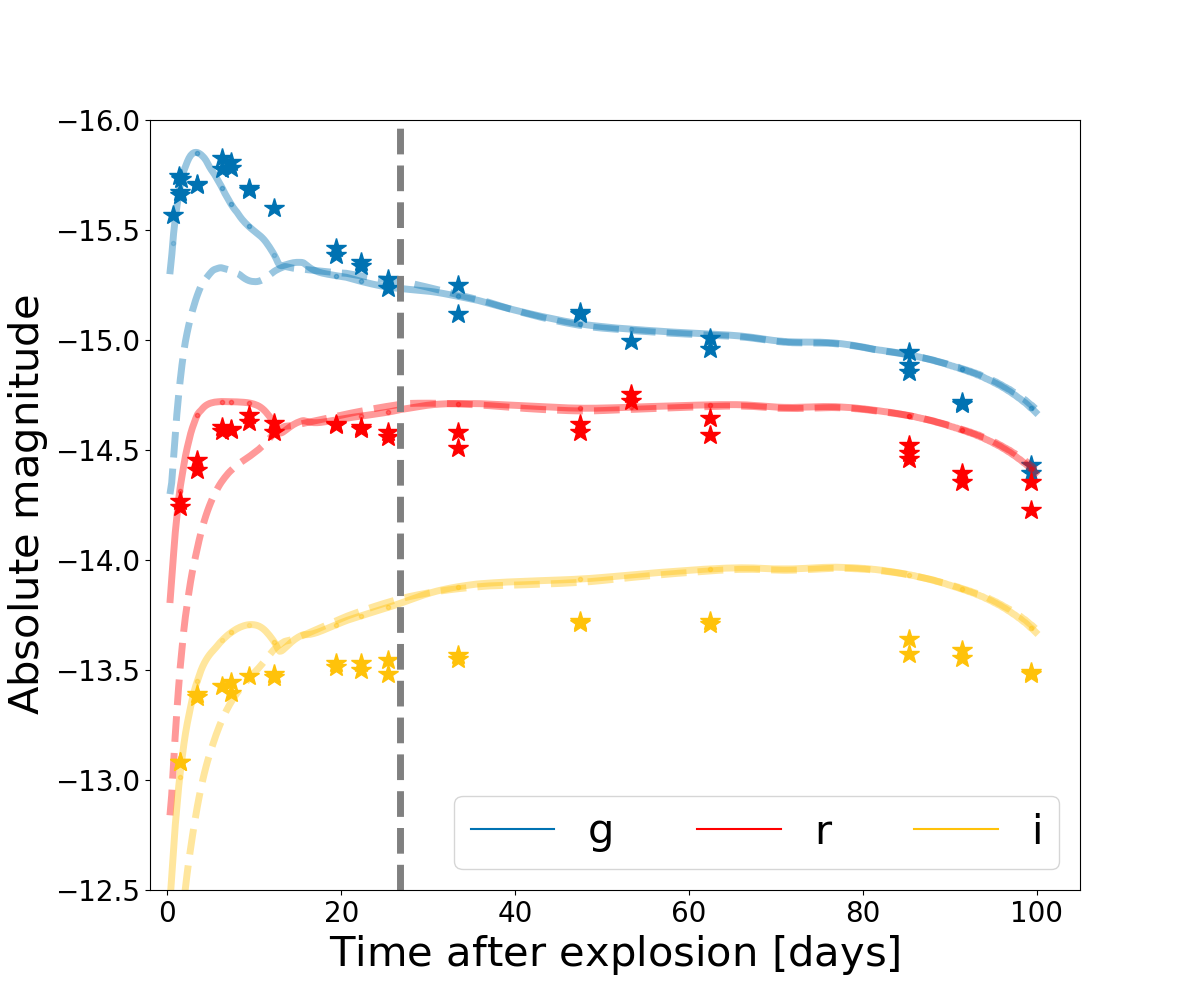}
    \caption{Best CSM model fit compared to SN~2021gmj $gri$ light curves. The best fit (solid lines) corresponds to a density of $K=2.4\times 10^{17} \ \mathrm{g} \ \mathrm{cm}^{-1}$ and a CSM radius extent of $R=750 \ R_{\odot}$. We also show the model without CSM using dashed lines. The vertical dashed line marks the end of the $s_1$ phase which is used as the earliest epoch in the progenitor mass-explosion energy fits and the latest epoch in the CSM parameter fit. We can see that adding CSM improves the fit significantly at early epochs.}
    \label{fig:CSMgri}
\end{figure}

The total CSM mass of our best-fit model is $M_{\rm CSM} = 0.025$ \msun
. This CSM mass is lower than those in the sample of \citet{Morozova2018} (without considering the lower limits), but it seems to follow the observed trends for a low mass and low energy explosion. Although the CSM mass is low, its importance for the early light curve is evident in Figure \ref{fig:CSMgri}, where we show the light curves of the models with and without CSM. The fit in the $g$ and $r$ bands improves significantly owing to the excess early emission coming from the ejecta-CSM interaction. However, the $i$-band model does not show a proper fit. This indicates that although the hydrodynamic modeling that includes CSM is an improvement, our simplified treatment of the model spectral energy distribution and our steady-wind assumption for the mass loss maybe not be a good representation of the density profile of the immediate vicinity of the progenitor of SN~2021gmj.  

\section{Spectroscopic Evolution}
\label{sec:spec_evo}
\subsection{Overall Evolution}
\begin{figure*}[t]
    \centering
    \includegraphics[width=2\columnwidth]{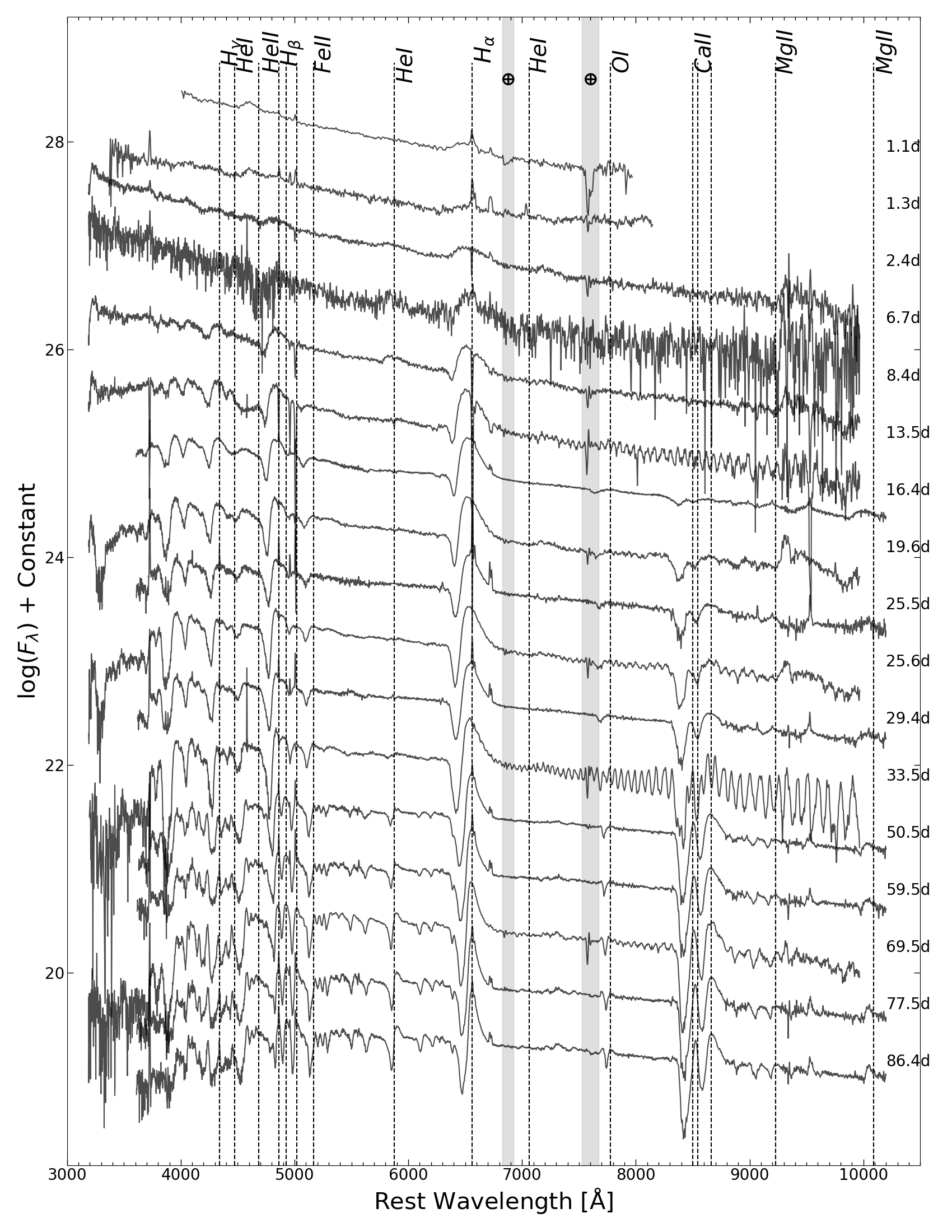}
    \caption{Optical spectral evolution of SN~2021gmj in the photospheric phase. The phase of each observation, in days since explosion, is labeled for each spectrum. The gray vertical bands with the $\oplus$ symbol mark the regions of strongest telluric absorption. The dashed vertical lines mark the rest wavelengths of Balmer lines of hydrogen along with other elements. The day 33.54 spectrum suffers from bad fringing at near-infrared wavelengths; fringing is also visible in the day 13.54 spectrum.}
    \label{fig:spec_evo}
\end{figure*}

The optical spectroscopic evolution of SN~2021gmj is shown in Figure \ref{fig:spec_evo}. The earliest spectra exhibit a blue continuum with broad H$\alpha$ emission and little absorption. These early-time spectra show no signs of prominent, narrow emission lines --- sometimes referred to as ``flash'' features --- often seen in SNe~II hours to a few days after explosion \citep[e.g.,][]{Bruch2021}. There is strong and broad emission around 4600~\AA\ instead. Further analysis of this emission in the early spectra and implications are discussed in Section \ref{sec:early_emission}.

\begin{figure}[t]
    \includegraphics[width=\columnwidth]{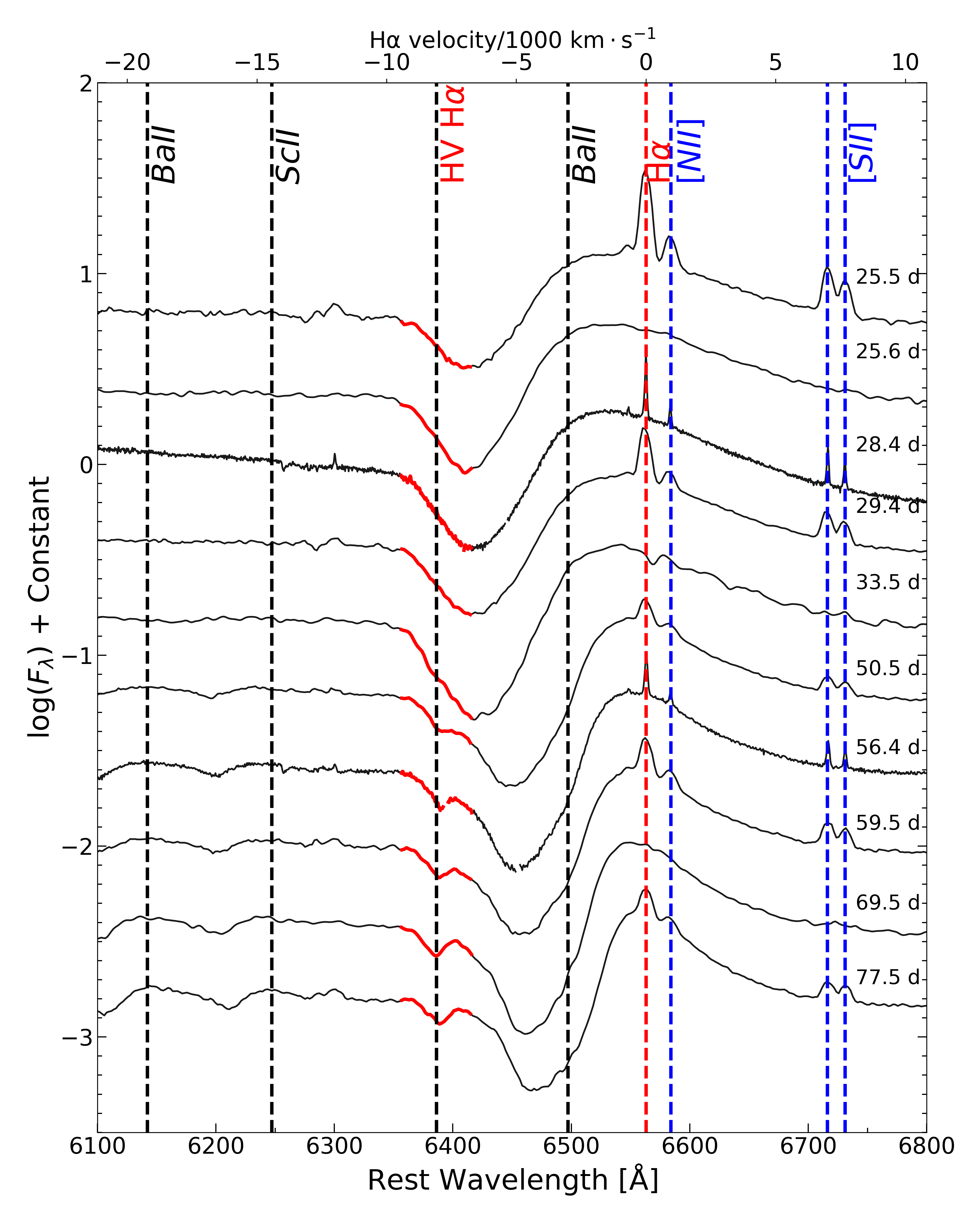}
    \caption{Spectral time series of SN~2021gmj showing the evolution near the H$\alpha$ line. We highlight the wavelength positions of \ion{Ba}{2} and \ion{Sc}{2} in black; we also label with blue font the wavelengths of host-galaxy nebular lines which contaminate the SN spectra. These narrow lines most likely originate from a nearby star-forming region. In red we mark H$\alpha$ together with the absorption component that we label as  high-velocity (HV) H$\alpha$ at $\sim 8000$ $\mathrm{km} \ \mathrm{s}^{-1}$.}
    \label{fig:cachito}
\end{figure}

Over time the SN becomes redder and the P~Cygni profiles of the hydrogen features become more prominent. Starting with the spectrum taken 13 days past explosion, \ion{Fe}{2} $\lambda 5169$ is observed and can be used to trace the photosphere. Around 20 days past explosion, more Fe-group lines appear, namely \ion{Fe}{2} $\lambda\lambda 5267$, 5363. At 20 days we also see the emergence of the \ion{Ca}{2} near-infrared (NIR) triplet around 8500~\AA. In Figure \ref{fig:cachito}, we show SN~2021gmj spectra in the region near H$\alpha$. Fifty days past explosion there is evidence of an absorption feature blueward of H$\alpha$. At this phase, this feature is identified as high-velocity H$\alpha$ and is attributed to a dense shell that forms due to the interaction of CSM with the SN ejecta \citep{Chugai2007,Gutierrez2014}. With this interpretation we measure a velocity of $\sim 8100$ $\mathrm{km} \ \mathrm{s}^{-1}$ for this feature. 

\begin{figure*}[t]
    \centering
    \includegraphics[width=2.0\columnwidth]{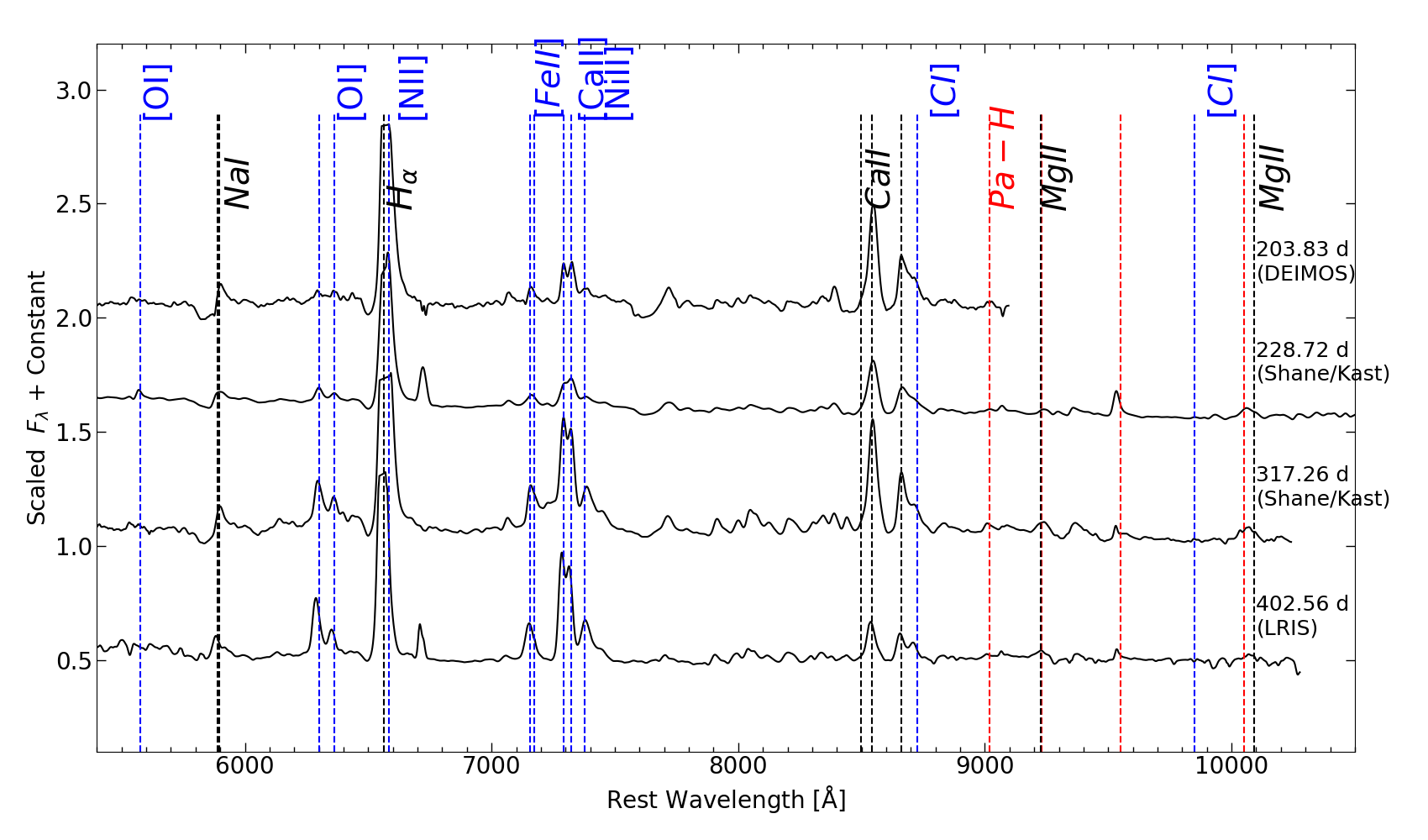}
    \caption{Spectroscopic sequence of SN~2021gmj in the nebular phase. In dashed lines we mark the wavelength of important allowed transitions (black), forbidden transitions (blue), and the Paschen series in the NIR (red). Near 6700 \AA \ there is likely an artifact coming from a bad subtraction of host emission from [\ion{S}{2}].}
    \label{fig:nebular}
\end{figure*}

After the fall from the plateau, we obtained three spectra at 204, 229, and 402 days past explosion, which we show in Figure \ref{fig:nebular}. They exhibit a clear transition from a P~Cygni-dominated spectrum to an emission-line nebular spectrum where we can observe the growing strength of forbidden lines like [\ion{O}{1}], [\ion{Fe}{2}], and [\ion{Ca}{2}]. The presence of nebular emission lines tells us that the expanding ejecta have become optically thin. This marks the end of the photospheric phase of the SN. These emission lines will be used in Section \ref{sec:neb} to constrain the progenitor mass and yields.  

\subsection{Metallicity of the SN}
\citet{Anderson2014} have shown that the  equivalent width of the \ion{Fe}{2} $\lambda5018$ line can be used to constrain the metallicity of SNe~II. In our spectra at 50 days we measured the \ion{Fe}{2} $\lambda5018$ line equivalent width; its value of $12$ \AA\ is closest to the 0.4 $Z_\odot$ measurement from the models used by \citet{Anderson2016}. This measurement is in agreement with the host oxygen abundance determinations of 0.4--0.5 solar described in Section \ref{sec:host}.

\subsection{Comparison with Other SNe~II}

It is well established that expansion velocities are correlated with luminosity for Type II SNe in general \citep{Hamuy2003,Kasen2009}. In particular, LL~SNe~II are expected to have low expansion velocities during the plateau, but there are counterexamples in the literature \citep{Dastidar2019,Rodriguez2020}. We now wish to establish a spectral comparison of SN~2021gmj with a range of SNe~II and see how well it fits to known correlations. 

\begin{figure*}[t]
    \centering
    \includegraphics[width=2.0\columnwidth]{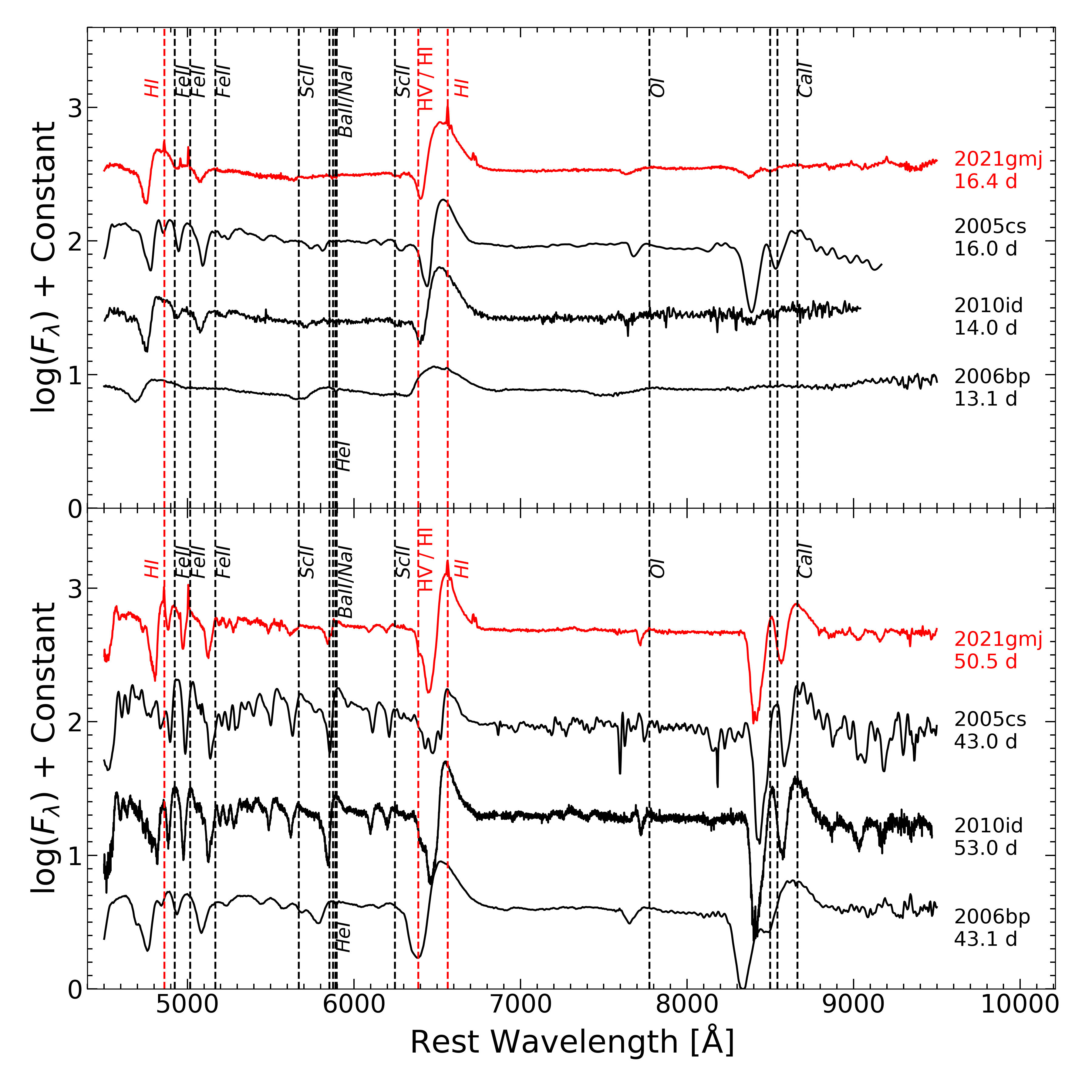}
    \caption{The upper panel shows a comparison of SN~2021gmj with other LL~SNe~II and SN~2006bp around 15 days past explosion. SN~2021gmj is most similar to SN~2010id, with weak and narrow H$\alpha$ absorption and missing blue lines. The lower panel displays a comparison with other LL~SNe~II around 50 days past explosion. The line profiles are narrow, consistent with what is expected from LL~SNe. SN~2006bp is included as an example of a more ordinary SN~II where the line profiles are broader. All the spectra have been normalized after a black-body fit. SN~2021gmj has intermediate spectral characteristics between the low-luminosity end of LL~SNe~II and more energetic Type II SNe.}
    \label{fig:spec_compare_comb}
\end{figure*}

Figure \ref{fig:spec_compare_comb} shows the spectrum of SN~2021gmj at 16 days past explosion compared to other LL~SNe~II at similar phases: SN~2005cs \citep{Pastorello2006} and SN~2010id \citep{Gal-Yam2011}. The epoch of 16 days was chosen to compare to the other objects at a very similar phase when the H$\alpha$ P~Cygni profiles are well developed. We also include SN~2006bp \citep{Quimby2007} for reference as a more luminous SN~II with broader lines. The shallow and narrow absorption of H$\alpha$ and H$\beta$ in the SN~2021gmj spectrum is most similar to SN~2010id. At this stage, many lines like \ion{Na}{1}~D and the \ion{Ca}{2} NIR triplet have not developed yet, unlike in the LL~SN~II 2005cs.

Figure \ref{fig:spec_compare_comb} also shows the day 50 spectrum of SN~2021gmj compared with the same sample at similar phases. The line profiles are narrow, consistent with what is expected from a LL~SNe~II. Metal lines also appear relatively shallow, consistent with the subsolar metallicity measurement. The lower velocity of the lines allows the triplet emission of \ion{Ca}{2} in the red part of the spectrum to be visible. We also observe the appearance of \ion{Ba}{2} and \ion{Sc}{2} lines, which are common in the colder spectra of LL~SNe~II \citep{Spiro2014,Pastorello2004}. Although SN~2021gmj shows narrower lines compared to SN~2006bp, the absorption lines are not as deep as in SN~2005cs, and the \ion{Ba}{2} $\lambda 6497$ line that distorts H$\alpha$ in SN~2005cs is not visible. This suggests that SN~2021gmj has intermediate spectral characteristics between the more extremely underluminous end of LL~SNe~II and brighter SNe~II.  

\subsection{Expansion Velocities}
\begin{figure}[h!]
    \centering
    \includegraphics[width=0.99\columnwidth]{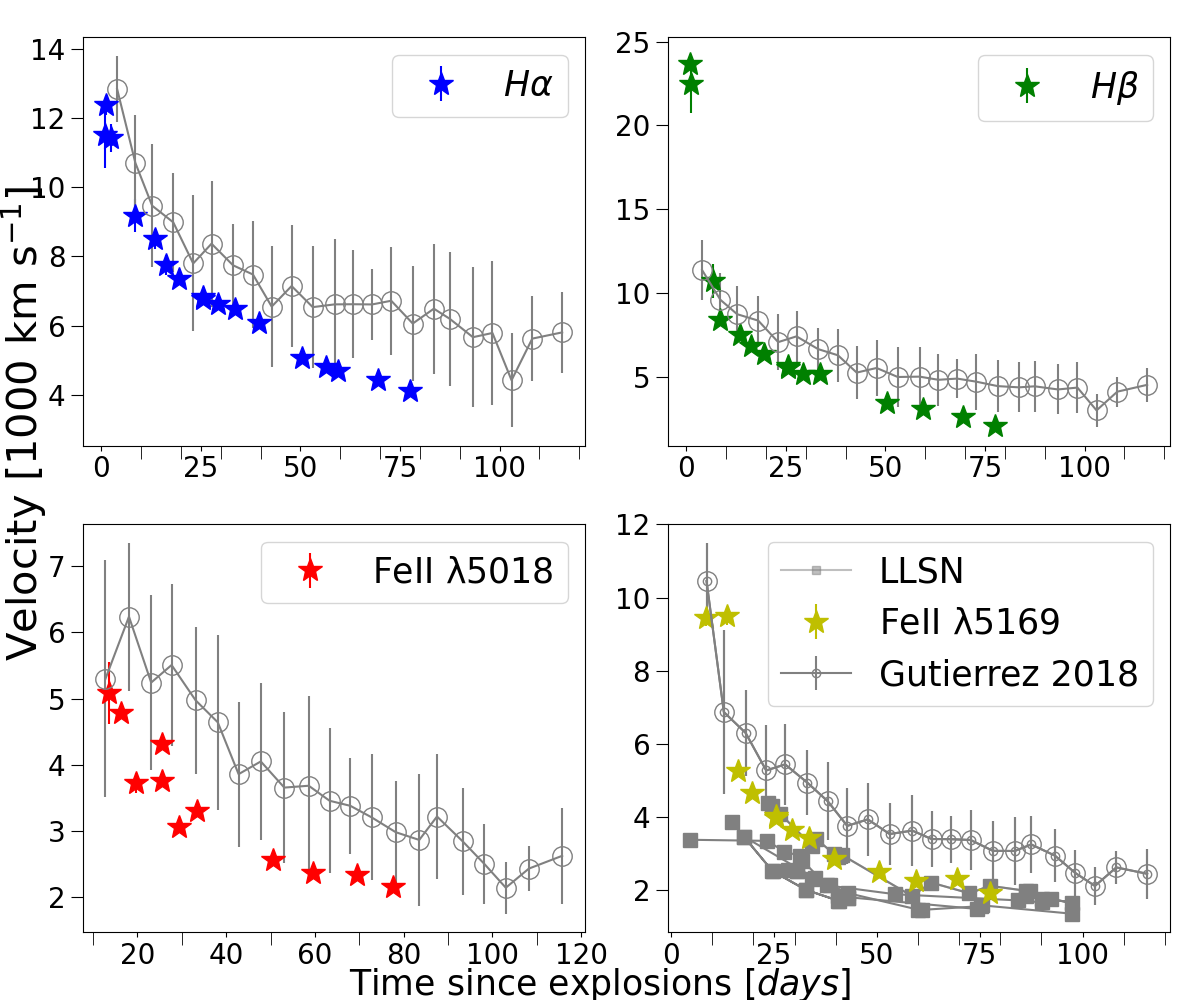}
    \caption{Expansion velocities as a function of phase for SN~2021gmj (star data points). The average and 1-sigma range values of a large sample of SNe~II from \citet{Gutierrez2017} are also shown (gray open circles). In the lower-right plot we include measured velocities of LL~SNe~II from the literature. SN~2021gmj exhibits below-average velocities with respect to the overall SNe~II population, but are in the higher range when compared to other LL~SNe~II.}
    \label{fig:velocity_evo}
\end{figure}

The expansion velocity evolution of SN~2021gmj is shown in Figure \ref{fig:velocity_evo} along with the average velocity values for a sample of SNe~II \citep{Gutierrez2017}. At early times, 5--10 days past explosion, SN~2021gmj has H$\alpha$ and H$\beta$ velocities of $\sim 8000$--10,000~\kms, slightly higher than normal for a LL~SNe~II \citep[][]{Pastorello2004,Spiro2014} and slightly lower than the average sample of \citet{Gutierrez2017}. The H$\alpha$ expansion velocities of SN~2021gmj decline more rapidly than the average sample to $\sim 7000$~\kms \ at 20 days past explosion, which is more consistent with what is seen in LL~SNe~II. In the lower-right panel of Figure \ref{fig:velocity_evo} comparing the \ion{Fe}{2} 5169 \AA \ velocities, we also include measured velocities from LL~SNe~II from the following literature: SN~1999br \citep{Pastorello2004}, SN~2002gd and SN~2003Z \citep{Spiro2014}, SN~2005cs \citep{Pastorello2009}, and SN~2009N \citep{Takats2014}. SN~2021gmj exhibits below-average velocities with respect to the overall SNe~II population, but in the higher range compared to other LL~SNe~II.

\subsection{Early-Time Spectra and Flash Signatures}
\label{sec:early_emission}

\begin{deluxetable*}{cccccc}
\tablecaption{SNe~II with a broad 4600 \AA\ feature in early-time spectra. \label{tab:llsne_ledge}
}
\tablecolumns{6}
\tablehead{
\colhead{Name} &
\colhead{Explosion date} &
\colhead{$z$} &
\colhead{Reference}
}
\startdata
SN 2002gd & 2452553.0 (4.0) & 0.007 & \citet{Anderson2014} \\
SN 2005cs & 2453549.5 (1.0) & 0.001 &  \citet{Pastorello2006,Silverman2017} \\
SN 2006bp & 2453834.5 ($^*$) & 0.0035 & \citet{Quimby2007} \\
SN 2010id & 2455454.5 (2.0) & 0.017 & \citet{GalYam2011} \\
ASASSN-14jb & 2456946.6 (3) & 0.006 & \citet{Meza2019} \\
SN 2018lab & 2458481.40 (1.0) & 0.00920 & \citet{Pearson2023} \\
\enddata
\tablecomments{Explosion dates and uncertainties are taken from the listed references. ($^*$) No uncertainty in explosion epoch given by \citet{Quimby2007}.}
\end{deluxetable*}

\begin{figure}[h!]
    \centering
    \includegraphics[width=\columnwidth]{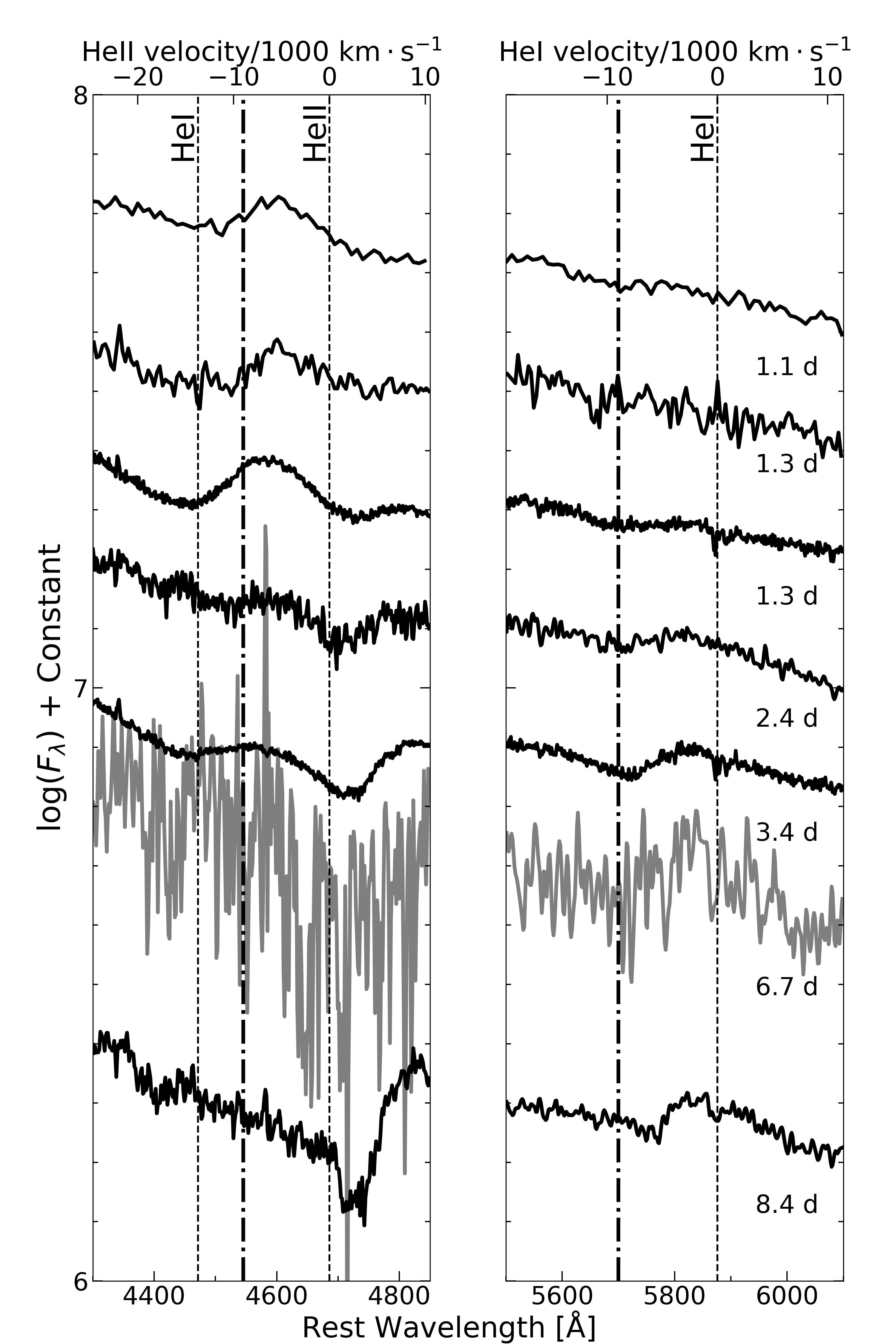}
    \caption{Comparison of the early-time blue emission, likely due to \ion{He}{2} $\lambda$4686 (left panel), to that of \ion{He}{1} $\lambda$5876 (right panel). As  \ion{He}{2} becomes weaker (i.e., the He recombines), the \ion{He}{1} $\lambda$5876 feature becomes stronger.}
    \label{fig:early_emission}
\end{figure}

\begin{figure*}[t]
    \centering
    \includegraphics[width=0.85\textwidth]{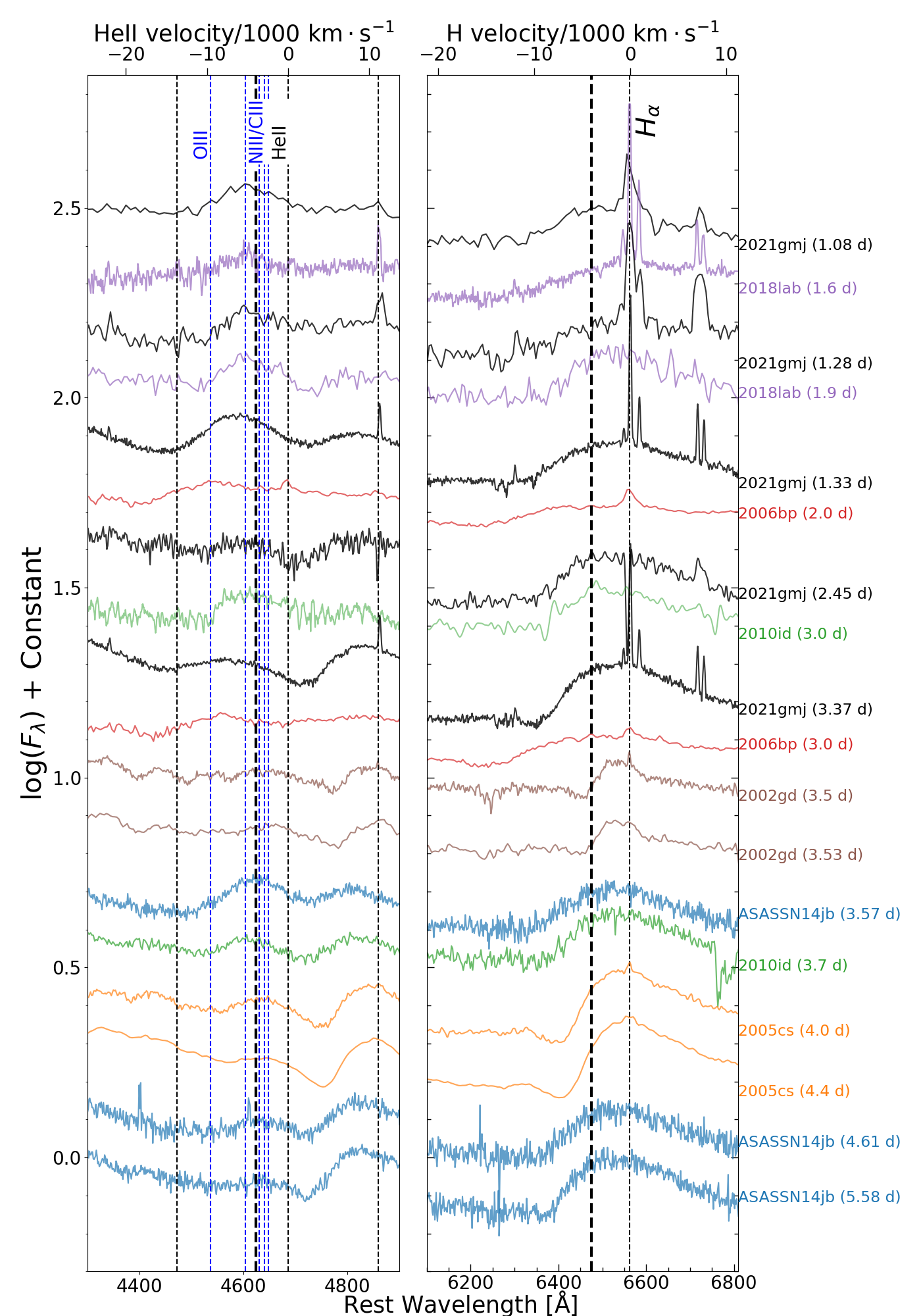}
    \caption{Comparison of the early emission in SN~2021gmj and other SNe with similar spectral profiles during the first 6 days after explosion, centered at \ion{He}{2} $\lambda$4686 (left panel) and at H$\alpha$ (right panel). In each panel the thick black dashed lines show the position of a blueshift of 4000 $\mathrm{km} \ \mathrm{s}^{-1}$ with respect to the rest wavelength of \ion{He}{2} $\lambda$4686 and H$\alpha$. Thin blue dashed vertical lines are drawn at the rest wavelengths of \ion{He}{2} $\lambda$4686, \ion{N}{3}, \ion{C}{3}, \ion{O}{3}, and H transitions. All  spectra have been normalized to a black-body fit. All SNe in the sample have emission near 4600 \AA, indicating high ionization of the ejecta at early times. Explosion epoch, redshift, and references for this comparison sample can be found in Table \ref{tab:llsne_ledge}.}
    \label{fig:heI-compare}
\end{figure*}

\begin{figure*}[t]
    \centering
    \includegraphics[width=0.8\textwidth]{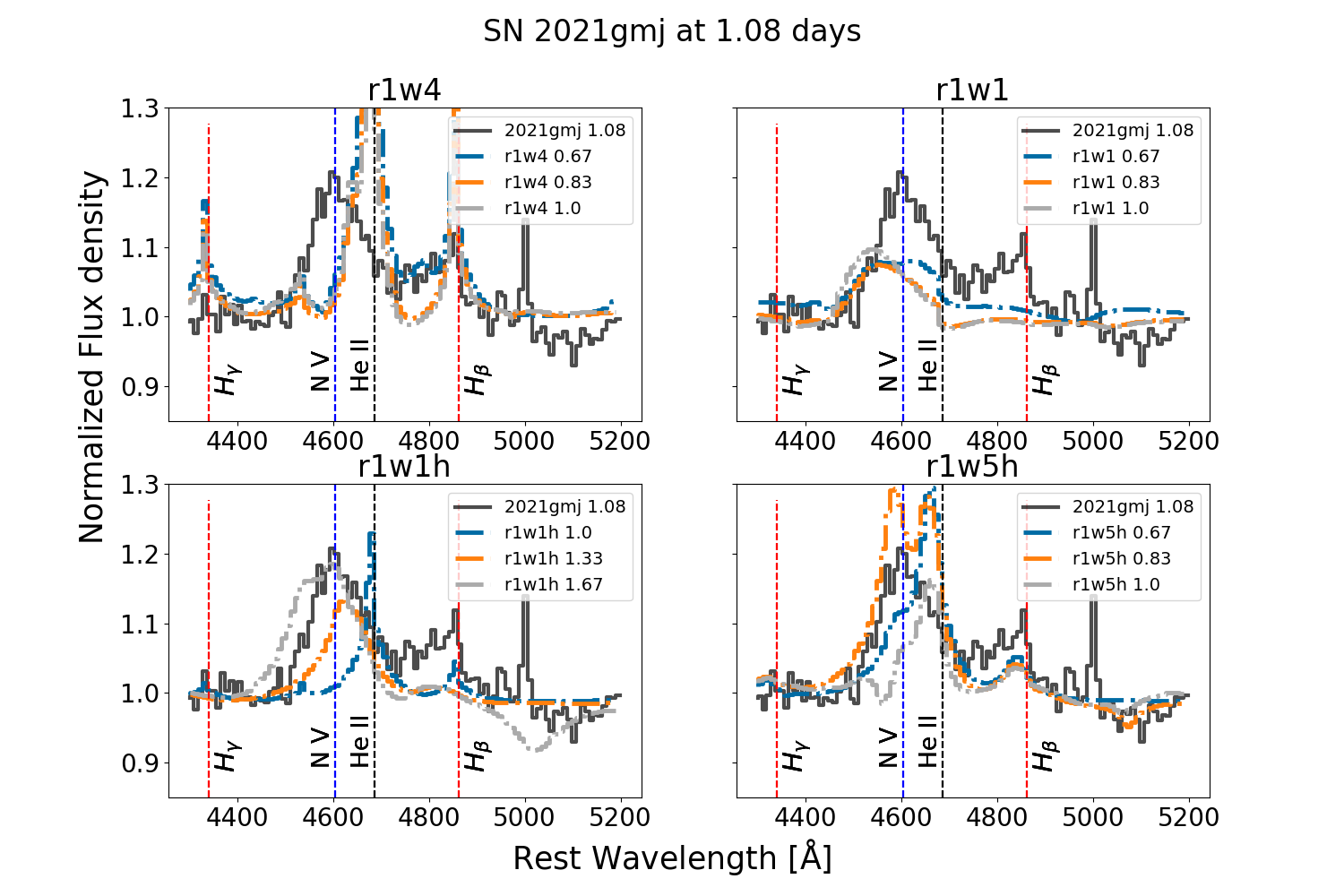}
    \caption{Comparison of SN~2021gmj first spectra around 4600 \AA\ with a selection of the models from \cite{Dessart2017}. Each panel compares to a given model at the three closest epochs. All spectra have been normalized to a black-body continuum fit. The models have been convolved to a Gaussian with the width of the resolution of our observation (18 \AA\ for the first LT spectra) and have been binned to match the observed resolution. In general, all models underestimate the H$\beta$ emission. SN~2021gmj is most similar to the extended atmosphere models, r1w1h and r1w5h, just  hours after the strong and narrow emission lines of \ion{He}{2} $\lambda 4686$ and \ion{N}{5} $\lambda 4604$ (black and blue dashed lines in the figure) evolve into a broader profile. }
    \label{fig:lucmodels}
\end{figure*}

As discussed in Section \ref{sec:spec_evo} and shown in Figure \ref{fig:early_emission}, the early spectra of SN~2021gmj have a broad emission feature around 4600~\AA. Similar features have been described as ``ledge-shaped," and may arise from a blend of high-ionization lines \citep[see, e.g.,][]{Bruch2021}. {Although in other works the ``ledge'' feature is reffered to as a flash feature, in this work we choose to call it the ledge to differentiate this broad emission with the narrow ones observed in very early spectra of SNe~II.}
Past work has also analyzed this feature in other SNe~II. \citet{Quimby2007} suggest that the ledge could be due to either \ion{He}{2} $\lambda$4686 or a \ion{N}{2} $\lambda\lambda$4480, 4630 blend. If this emission was caused by \ion{N}{2}, then \ion{N}{2} $\lambda\lambda$5490, 5680 should also be visible \citep{Dessart2005b}; however, this line is absent from our early spectra of SN~2021gmj. Similar to the argument with the \ion{N}{2} line, one can look for transition lines of other elements which can be comparable in brightness at high temperatures. In the case of C and N this is challenging because there are strong nearby lines such as H$\delta$ or \ion{He}{1}\footnote{In the case of \ion{N}{5} (in LTE conditions), one expects \ion{N}{4} emission near 4100 \AA \ at $k_B T = 1.6$--1.7 eV. In the case of \ion{C}{3}, there is a strong \ion{C}{4} line very close to 5800 \AA \ at a temperature of $k_B T \approx 1.2$ eV. We do not observe any of these transitions.}. 

Following \citet{Quimby2007}, as the ejecta cool and He recombines, the \ion{He}{2} $\lambda4686$ would become depopulated while \ion{He}{1} $\lambda5876$ would grow stronger. This is precisely what we see in SN~2021gmj (Figure \ref{fig:early_emission}), suggesting that the early-time emission is due to blueshifted \ion{He}{2} $\lambda4686$ and that the ejecta are still highly ionized at these phases. If this feature is associated with \ion{He}{2}, it has a velocity of $4000\pm100\ \mathrm{km} \ \mathrm{s}^{-1}$ as measured by fitting a Gaussian and a linear continuum in the region. This value is not unreasonable compared with other SNe.
In SNe~II, P~Cygni emission maxima are expected to be blueshifted at early times by as much as 5000 $\mathrm{km} \ \mathrm{s}^{-1}$ \citep[][]{Anderson2014b,Dessart2005b}. However, the amount of blueshift is correlated with the slope of the $V$-band light curve at 30 days \citet{Anderson2014b}.
There we see that flatter light curves have smaller blueshifts. 
This would imply a blueshift of $\sim 1000\mathrm{km} \ \mathrm{s}^{-1}$ for the peak of the P~Cygni profiles in the spectra of SN~2021gmj 30 days after explosion; extrapolating to early phases, we expect a blueshift of $\sim 2000$--3000 $\mathrm{km} \ \mathrm{s}^{-1}$. Indeed, looking at H$\alpha$ in SN~2021gmj, the emission appears blueshifted by no more than 3000 $\mathrm{km} \ \mathrm{s}^{-1}$. Thus, it seems unlikely that the \ion{He}{2} $\lambda$4686 blueshift would be as high as 4000 $\mathrm{km} \ \mathrm{s}^{-1}$.

To place SN~2021gmj in context, we compare the early-time spectra with spectra from a sample of LL~SNe~II that show emission near \ion{He}{2} $\lambda 4686$ (Figure \ref{fig:heI-compare}). We also included the normal-luminosity SN~2006bp, which was one of the first SNe~II with very early spectra showing flash features. Explosion epoch, redshift, and references for this comparison sample can be found in Table \ref{tab:llsne_ledge}. In the LL~SNe~II, the ledge feature peaks near the high-ionization lines of \ion{C}{3}, \ion{N}{3}, and possibly \ion{N}{5} $\lambda4604$ \AA. SN~2006bp, however, is brighter and has broader features than the other LL~SNe~II, having the peak near \ion{O}{3} lines. We observe that no supernova shows a feature with a peak near the rest frame of \ion{He}{2} $\lambda4686$. A relevant observation is that the blueshift of the ledge feature, if we interpret it as \ion{He}{2} $\lambda4686$, is inconsistent with the blueshift of H$\alpha$ in all of the spectra shown. Despite their mismatched velocities, this feature could still be interpreted as \ion{He}{2} $\lambda4686$ if there is an ionization stratification of the ejecta. If higher velocity material is at higher ionization than the rest of the slower ejecta, the ionized \ion{He}{2} lines may be observed at higher velocity than the Balmer lines. The higher ionization at higher velocities can occur as CSM interaction of the high-velocity ejecta is exposed to the hard radiation from the ejecta-CSM shock region. A similar argument is found in the recent work of \citet{Chugai2024}. The authors propose that the \ion{He}{2} $\lambda4686$ emission comes from dense fragments embedded in the forward-shock region that undergo Rayleigh-Taylor instabilities. We conclude that while the ledge feature seems common in LL~SNe~II \citep{Pearson2023}, we cannot definitively identify its origin.

To further understand the origin of the early emission,  we compare the first spectrum of SN~2021gmj with the model spectra from \citet{Dessart2017} (Figure \ref{fig:lucmodels}). These models explode 15 \msun\ RSG progenitors with different density structures, controlled by differing mass-loss rates and atmospheric scale heights. The r1w1 and r2w1 models have progenitor radii of $R_{*}=501$ \rsun \ and $R_{*}=1107$ \rsun\ (respectively) and a mass-loss rate of $\dot{M} = 10^{-6}$ \msun $\mathrm{yr}^{-1}$ for both. Model r1w1h has an extended atmospheric density scale height of $0.3\,R_{*}$ up to a density of $10^{-12} \ \mathrm{g} \ \mathrm{cm}^{-3}$ followed by a power law with exponent 12 until it reaches the density profile of a $\dot{M} = 10^{-6}$ \msun $\mathrm{yr}^{-1}$ wind at a radius of $\sim 2\times10^{14}$ cm. Finally, model r1w5h is similar to r1w1h but with a scale height of $0.1\,R_{*}$, which then decreases to a wind of $\dot{M} = 5 \times 10^{-3}$ \msun $\mathrm{yr}^{-1}$ and finally to a wind of $\dot{M} = 10^{-6}$ \msun $\mathrm{yr}^{-1}$ at a radius of $\sim 2\times10^{14}$ cm. 

These two extended-envelope models (r1w1h and r1w5h) attempt to represent the complex extended atmospheres observed in RSGs, which can have inflows/outflows and inhomogeneities up to a few stellar radii \citep{Arroyo2015,Kervella2016,Humphreys2022,Gonzalez2022,Goldberg2022}. To do the comparison we first convolve the model spectra with a Gaussian with a width equal to the instrumental resolution of the SN~2021gmj spectra, and bin the models to the  resolution of the observations. We then fit a black body to the continuum of both the model and our observed spectra and normalize each spectrum. Like SN~2018lab \citep{Pearson2023}, SN~2021gmj is better fit by the extended-atmosphere models (r1w1h and r1w5h). 
In the r1w1h and r1w5h models, the initially strong and narrrow \ion{He}{2} $\lambda 4686$ and \ion{N}{5} $\lambda 4604$ emission profiles blend together to create a broader profile centered around 4600 \AA\ which is similar to the 4600 \AA\ feature seen in SN~2021gmj.

Following \citet{Dessart2017}, the broad diversity of the emission in early-time spectra can be explained by the origin of the broadening of these high-ionization lines. If the line of a given element is formed in a slow wind of unshocked material the line profile is symmetric, driven by incoherent electron scattering. Later, when the dense wind is swept up by the supernova ejecta, a dense shell forms that will have as a spectral profile a Doppler broadened and blueshifted emission peak \citep{Dessart2017}. This scenario seems consistent with the early observations of LL~SNe~II. Therefore, although we cannot rule out the high-velocity \ion{He}{2} origin of this feature,  the spectral evolution is consistent with a progenitor exploding into a higher density material. 

\section{Progenitor Mass through Nebular Spectral Analysis}
\label{sec:neb}

As the plateau phase ends, the photosphere recedes from the hydrogen-rich outer layers and we are able to observe emission from the core. This allows us to constrain the elemental composition of the inner layers of the ejecta and the progenitor mass. In particular, the emission from the forbidden lines of [\ion{O}{1}] is a good probe of the oxygen mass in the ejecta and therefore to the main-sequence mass of the star \citep{Woosley1995}. We use the nebular-spectra models from \citet[][J12 models hereafter]{Jerkstrand2012} to constrain the progenitor mass of SN~2021gmj. These models explode 12, 15, and 19 \msun \ single-star progenitors with a piston, giving a total of 1.2 foe of kinetic energy, and fix the mass of the Fe/He zone to give a nickel mass of 0.062 \msun.

\begin{figure*}[t]
    \centering
    \includegraphics[width=2.0\columnwidth]{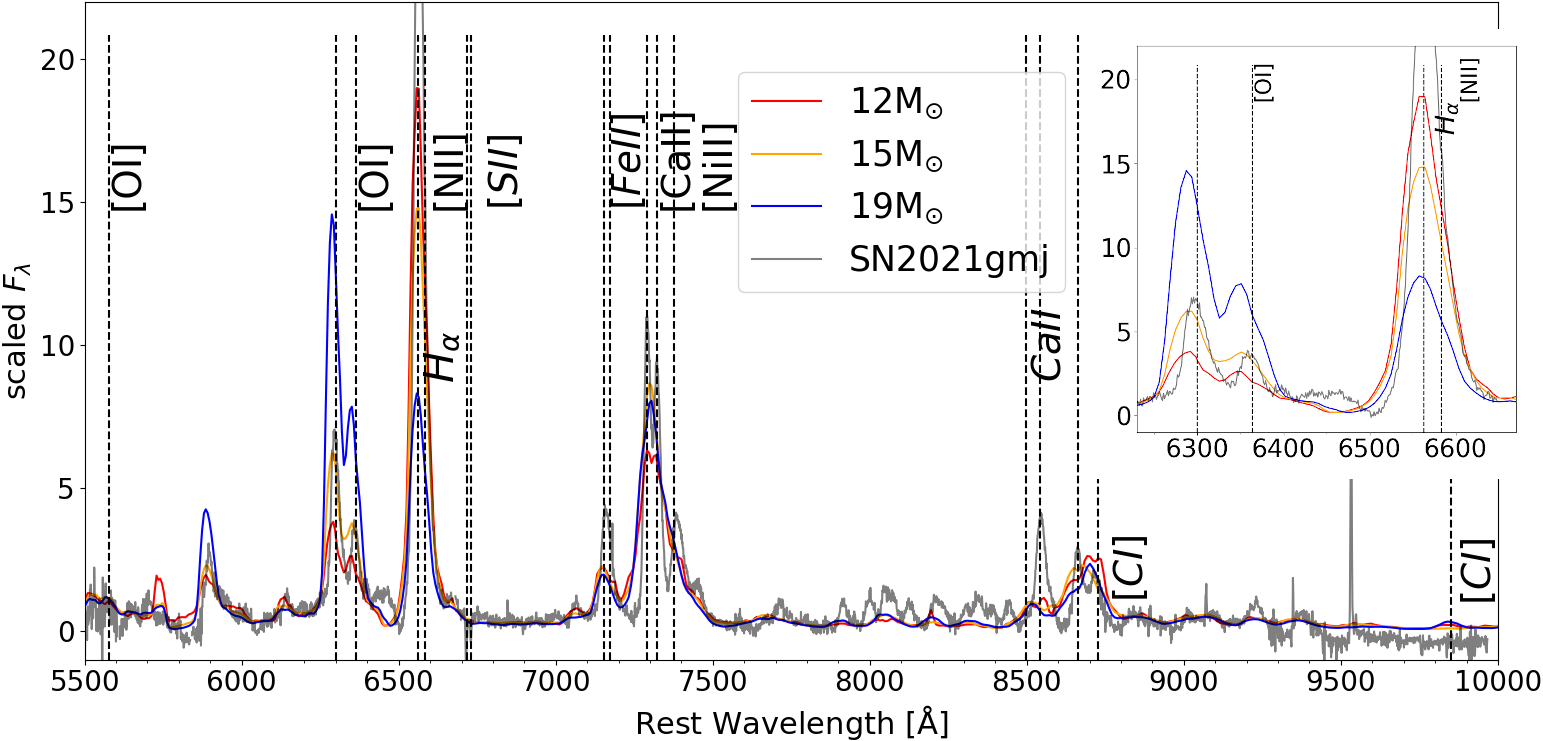}
    \caption{Spectra of SN~2021gmj (dereddened) at 402 days past explosion compared with the 12 \msun, 15 \msun,  and 19 \msun\ models from \citet{Jerkstrand2012} at 400 days. Both the observed and model spectra have been normalized to the integrated flux in the 5500--10,000 \AA\ region. On the upper right we display an inset of the spectra around [\ion{O}{1}] and H$\alpha$. Our analysis shows that the relative flux of the [\ion{O}{1}] doublet to both the [\ion{Ca}{2}] flux and to the cobalt-decay power is very close to the 12 \msun\ model.} 
    \label{fig:nebular_models}
\end{figure*}

In Figure \ref{fig:nebular_models}, we compare our Keck/LRIS observation at $\sim400$ days past explosion with the J12 models. To identify which model best characterizes the overall spectrum of SN~2021gmj, we normalize the model spectra to the total flux over the observed wavelength range of  SN~2021gmj (5500--10,000 \AA). This has the effect of aligning the continuum regions, allowing us to compare the observed and model spectra. Overall, the models have broader lines than SN~2021gmj. The width of the lines is related to the velocity of the inner layers, which correlates with the explosion energy. Additionally, the small line width allows us to distinguish the [\ion{Ni}{2}] $\lambda7378$ and [\ion{C}{1}] $\lambda8727$ emission clearly. We can therefore conclude that SN~2021gmj has a lower explosion energy than these models,  consistent with our SNEC modeling. 

Next we consider in detail the emission of specific lines and discuss the progenitor mass. As mentioned, the emission from [\ion{O}{1}] is related to the progenitor mass. We measured the integrated flux of the [\ion{O}{1}] doublet by fitting Gaussian profiles over a flat continuum for SN~2021gmj and the model spectra. The luminosity (or flux) is then normalized by the cobalt decay power at 400 days to obtain the fraction of oxygen luminosity relative to cobalt decay. This fraction is 0.22 for SN~2021gmj and is equivalent to the model of 12 \msun\  within uncertainties ($\sim 5$--10\% assuming that the distance is the main source of statistical uncertainty). 

Some authors prefer to measure the ratio of [\ion{O}{1}] to [\ion{Ca}{2}] as this can minimize the effects of extinction, distance, and cobalt decay, allowing a better comparison of supernovae to each other \citep{Elmhamdi2011,Hanin2015,Fang2018,Fang2019,Hiramatsu2021a,Hiramatsu2021b,Fang2022}. We estimate the ratio for SN~2021gmj, measuring the flux of the [\ion{Ca}{2}] $\lambda\lambda 7291$, 7323 doublet as we did with [\ion{O}{1}]. In the case of [\ion{Ca}{2}], to avoid overestimating the flux we also include in our fit the visible neighboring lines [\ion{Fe}{2}] $\lambda\lambda 7155$, 7453 and [\ion{Ni}{2}] $\lambda\lambda 7378, 7412$ as extra gaussians profiles to the fitting model. We measure J12 model line ratios following the same procedure as well. The measured luminosity ratio [\ion{O}{1}]/[\ion{Ca}{2}] = 1/2 corresponds to the value of the 12 \msun \ model. 
 
 A relevant consideration when comparing SN~2021gmj to these nebular models is the mismatch in nickel mass, where the model has a factor of $\sim 4.5$ more nickel than SN~2021gmj itself. While the empirical scaling accounts for the difference in flux due to the differences in Ni mass, the increasing nickel mass in the models increases the ionization and electron density, which can suppress forbidden-line emission. In particular, Ca and O forbidden lines can be relatively weaker for higher nickel mass values \citep{Dessart2020}. We expect that our measurement using the ratio of these lines can alleviate at least partially this uncertainty. 

\citet{Dessart2020} observe that [\ion{Ca}{2}] can be a very efficient coolant, such that an increased Ca mass fraction in the oxygen shell will cause the [\ion{O}{1}] emission to be quenched. This quenching would cause us to underestimate the progenitor mass. However, the mixing of O and Ca shells is expected to occur in higher progenitor masses ($\gtrsim 17$ \msun); as we find that the progenitor mass of SN~2021gmj is well below 17 \msun, this is unlikely to affect our results. All our analyses assume that the progenitor of SN~2021gmj is a single star and that the mass-loss history of the progenitor is normal when compared to the assumed expected values used in \citet{Jerkstrand2012} (\citet{Niewenhujizen1990} parametrization is used in that work).
 
Our analysis allows us to conclude that SN~2021gmj has a progenitor mass of $M_{\rm ZAMS} \approx 12$ \msun, and nickel mass and explosion energy compatible with a less massive progenitor. This result is further supported by the hydrodynamic modeling of the light curves, which gave a progenitor mass of 10 \msun\ and an explosion energy of 0.294 foe. 

\section{Discussion}
\label{sec:analysis}

In Section \ref{sec:earlylc}, our light curve modeling suggests the presence of CSM around the progenitor of SN~2021gmj. Here we will analyze additional evidence supporting or disputing the presence of CSM interaction for SN~2021gmj.

There are two spectroscopic observations that may support the CSM scenario: the HV H$\alpha$ in the plateau spectra and the early emission near 4600 \AA. 
The HV H$\alpha$ can be explained by the presence of a wind. As discussed by \citet{Chugai2007}, either the unshocked wind or the cold dense shell that forms in the ejecta/wind interaction will have an increased opacity in material at high velocities due to the X-ray emission originating in the reverse and forward shocks. 

Coming back to the early emission, our analysis in Section \ref{sec:early_emission} points to either the presence of high-velocity \ion{He}{2}  and/or high-ionization lines of C and N. Similar conclusions were obtained for other SNe~II.
The study of SN~2018fif \citep{Soumagnac2020} showed very early, low-resolution, spectra with hints of emission of \ion{He}{2}, \ion{N}{3}-\ion{N}{5}, and \ion{C}{4}. Similarly, SN~2016bkv \citep{Hosseinzadeh2018} exhibited distinct double-peaked emission from \ion{C}{3}/\ion{N}{3} and \ion{He}{2} up to 5 days after explosion. Finally, we mention SN~2021yja \citep{Hosseinzadeh2022,Vasylyev2022}, which also showed a multicomponent, ledge-like, feature. Although SN~2021yja was in the bright class of SNe~II, the progenitor mass constraint from pre-explosion imaging put an upper limit of 9 \msun, which is closer to the progenitor mass of LL~SNe~II. As in the case of SN~2021gmj, the early emission of both SN~2018lab and SN~2021yja was most similar to the extended-atmosphere progenitor models from \citet{Dessart2017}. This may be evidence that LL~SNe~II  show spectral signatures of interaction in the very first days after explosion. The interaction may arise from the SN shock ramming through the low-density atmosphere of the RSG. Thereafter, the shock may encounter a wind of lower density which may extend the duration of the spectral signatures. The ejecta-atmosphere/CSM interaction may increase the peak brightness, like in the case of SN~2021yja. Future studies of early interaction signatures would benefit from higher-resolution spectra, as the current resolution does not allow a definite answer regarding the origin of the 4600 \AA\ line; it is possible that we may be observing other element transitions besides ionized He. 

Both the light curve modeling and the extended-star spectral models require a density of $\sim10^{-10}$--$10^{-11} \mathrm{g\ \mathrm{cm}^{-3}}$, a distance of just 1.5 times the progenitor-star radius. Whether this mass is associated with a high-density wind originating from a sudden increase in mass loss very near in time to the explosion \citep{Quataert2012,Smith2014,Fuller2017,Kuriyama2020}, or whether the material is just an extended atmosphere \citep[e.g.,][]{Mcley2014,Soker2021}, is something to study in future work.

\citet{Murai2024} have recently published a study of SN~2021gmj. Their analysis greatly coincides with ours, obtaining consistent peak brightness, expansion velocities, progenitor mass and ejected nickel mass. The progenitor mass of 12 \msun \ was obtained through a nebular spectral analysis similar to ours. They also used hydrodynamic models of progenitors with CSM, and obtained a mass-loss rate of $\sim 10^{-3} - 10^{-2.5} M_{\odot}\cdot yr^{-1}$ with a progenitor of fixed mass 12 \msun. This mass loss rate also agrees with our measurements. If we assume a terminal wind velocity of 10 km/s like in \citet{Murai2024}, we obtain a mass-loss rate of $\sim 3 \cdot 10^{-3} M_{\odot}\cdot yr^{-1}$. Our progenitor mass is potentially lower than the mass obtained by \citet{Murai2024}. We trust the value from hydrodynamic simulations more than the one from the nebular analysis (12 \msun), so we did not fix the progenitor mass in our SNEC modeling (10 \msun). Both works do agree that the progenitor mass is not extremely low compared to more lower-luminous events like SN~2005cs \citep{Maund2005}. SN~2021gmj then, taking both studies, presents itself as moderate explosion in a moderate-mass RSG progenitor, very close to the brightness limit of the class of LL~SNe~II. This is in agreement with our current understanding of the progenitors of LL~SNe~II \citep{Spiro2014}. The luminosity of a SNe~II is expected to depend on the progenitor mass, while the early peak luminosity depends more directly on the progenitor radius \citep[e.g.][]{Young2004,Rabinak2011} and the amount of CSM \citep[e.g.][]{Morozova2018}. Thus observations of LL~SNe~II like SN~2021gmj, which show signs of CSM interaction, provide key information on the diversity of RSG explosions.

\section{Conclusions}

We have presented photometric and spectral observations of the Type II SN~2021gmj. It photometrically belongs to the class of LL~SNe~II, with a maximum brightness of $M_V = -15.45$ mag. SN~2021gmj presents a decline of $<0.01$ mag per 100 days on the plateau, which lasts about 100 days. Both the photometric and spectral comparison show that SN~2021gmj is similar to a sample of LL~SNe~II, including expansion velocities and bolometric luminosity. SN~2021gmj synthesized a moderate amount of radioactive nickel, \Ni $= 0.014 \pm 0.001$ \msun, which together with the low expansion velocities point to a moderate progenitor mass and explosion energy \citep{Hamuy2003,Sukhbold2016}. This is confirmed through our analysis of the nebular emission lines; the line-intensity ratios are consistent with the 12 \msun\ progenitor model. Overall the basic properties of SN~2021gmj derived in this work are consistent with the higher-luminosity end of LL~SNe~II and are in agreement with the continuous distribution of parameters of RSG explosions \citep{Spiro2014,Valenti2016}. 

The early light curve modeling favors the presence of CSM very close to the star, which translates to a small CSM mass in our simple steady wind assumptions. The SNEC best fit gives a CSM mass of $M_{\rm CSM}=0.025$ \msun\ over the progenitor of 10 \msun\ exploded with a moderate energy of $E=0.294$ foe.
The low CSM mass agrees with the early-time spectral features, which do not show any narrow emission with electron-scattering wings. Instead, we observe the presence of a probable blend of high-ionization emission lines of C, N, and high-velocity \ion{He}{2}. This indicates that the line-formation region is partially or completely swept up by the fast-moving ejecta so that the overall profile is similar to a broad and blueshifted P~Cygni line. This is further supported by the similarity of a spectroscopic emission feature around 4600 \AA\ in SN~2021gmj to radiative hydrodynamic spectroscopic simulations of extended stars with CSM. The presence of this broad feature in several LL~SNe~II suggests that interaction signatures may be common in this class.

Although an abundance of photometric and spectroscopic evidence exists for the presences of compact and low-density material surrounding the RSG progenitor of SN 2021gmj, further research is required to understand if LL~SNe~II with CSM signatures arise from extended RSGs or from RSGs that have an increase of mass loss in the final stages of stellar evolution. 

\label{sec:conclusions}

\section*{Acknowlegdements}
Time-domain research by the University of Arizona team and D.J.S.\ is supported by National Aeronautics and Space Administration (NASA) grant 80NSSC22K0167, National Science Foundation (NSF) grants AST-1821987, 1813466, 1908972, 2108032, and 2308181, and the Heising-Simons Foundation under grant \#2020-1864.
J.E.A.\ is supported by the international Gemini Observatory, a program of NSF's NOIRLab, which is managed by the Association of Universities for Research in Astronomy (AURA) under a cooperative agreement with the NSF, on behalf of the Gemini partnership of Argentina, Brazil, Canada, Chile, the Republic of Korea, and the United States of America. Research by Y.D., N.M., and S.V.\ is supported by NSF grants AST-1813176 and AST-2008108.
The Las Cumbres group is supported by NSF grants AST-1911151 and AST-1911225.
K.A.B. is supported by an LSSTC Catalyst Fellowship; this publication was thus made possible through the support of grant 62192 from the John Templeton Foundation to LSSTC. The opinions expressed in this publication are those of the authors and do not necessarily reflect the views of LSSTC or the John Templeton Foundation.

L.G. acknowledges financial support from the Spanish Ministerio de Ciencia e Innovaci\'on (MCIN), the Agencia Estatal de Investigaci\'on (AEI) 10.13039/501100011033, and the European Social Fund (ESF) ``Investing in your future'' under the 2019 Ram\'on y Cajal program RYC2019-027683-I and the PID2020-115253GA-I00 HOSTFLOWS project, from Centro Superior de Investigaciones Cient\'ificas (CSIC) under the PIE project 20215AT016, and the program Unidad de Excelencia Mar\'ia de Maeztu CEX2020-001058-M.
The Las Cumbres group is supported by NSF grants AST-1911151 and AST-1911225.
A.V.F.'s group is supported by the Christopher R. Redlich Fund, Alan Eustace (W.Z. is a Eustace Specialist in Astronomy), Briggs and Kathleen Wood (T.G.B. is a Wood Specialist in Astronomy), Gary and Cynthia Bengier, Clark and Sharon Winslow, Sanford Robertson (Y.Y. was a
Bengier-Winslow-Robertson Postdoctoral Fellow), and many other donors.

We appreciate the expert assistance of the staffs at the observatories where data were obtained.  
Some of the data presented herein were obtained at the
W. M. Keck Observatory, which is operated as a scientific partnership among the California Institute of Technology, the
University of California, and NASA. The Observatory was made possible by
the generous financial support of the W. M. Keck Foundation.
A major upgrade of the Kast spectrograph on the Shane 3~m telescope at Lick Observatory, led by Brad Holden, was made possible through gifts from the Heising-Simons Foundation, William and Marina Kast, and the University of California Observatories. 
Research at Lick Observatory is partially supported by a generous gift from Google
This paper is based on observations made with the MuSCAT3 instrument, developed by the Astrobiology Center and under financial supports by JSPS KAKENHI (JP18H05439) and JST PRESTO (JPMJPR1775), at Faulkes Telescope North on Maui, HI, operated by the Las Cumbres Observatory.
%
Some observations reported here were obtained at the MMT Observatory, a joint facility of the University of Arizona and the Smithsonian Institution.

This work used the Weizmann Interactive Supernova Data Repository (WISeREP) \citep[\url{https://wiserep.weizmann.ac.il},][]{WISeREP}.
%
This research has made use of the NASA/IPAC Extragalactic
Database (NED), which is operated by the Jet Propulsion Laboratory, California Institute of Technology, under contract with NASA.
%
This work was  performed in part at Aspen Center for Physics, which is supported by National Science Foundation grant PHY-2210452.

\facilities{ADS, MMT (Binospec), Las Cumbres Observatory (Sinistro, FLOYDS, MuSCAT3), Keck
:I (LRIS), Keck: II (DEIMOS),Ekar:1.82 m (Afosc), Shane (KAST), Sleaford Observatory:Prompt, Swift (UVOT), NED, WISeREP.}

\software{  
astropy \citep{astropy:2013, astropy:2018}, corner \citep{corner}, emcee \citep{emcee}, FLOYDS pipeline \citep{FLOYDS}, \texttt{HOTPANTS} \citep{hotpants}, \texttt{lcogtsnpipe} \citep{Valenti2016}, MatPLOTLIB \citep{mpl}, NumPy \citep{numpy}, Scipy \citep{scipy}, \texttt{SNEC} \citep{Morozova2015}.
}

\bibliography{2021gmj}{}
\bibliographystyle{aasjournal}

\end{document}

%% file: affiliations.tex
\newcommand{\LCO}{\affiliation{Las Cumbres Observatory, 6740 Cortona Drive, Suite 102, Goleta, CA 93117-5575, USA}}
\newcommand{\UCSB}{\affiliation{Department of Physics, University of California, Santa Barbara, CA 93106-9530, USA}}
\newcommand{\KITP}{\affiliation{Kavli Institute for Theoretical Physics, University of California, Santa Barbara, CA 93106-4030, USA}}
\newcommand{\UCD}{\affiliation{Department of Physics and Astronomy, University of California, Davis, 1 Shields Avenue, Davis, CA 95616-5270, USA}}
\newcommand{\WIS}{\affiliation{Department of Particle Physics and Astrophysics, Weizmann Institute of Science, 76100 Rehovot, Israel}}
\newcommand{\OKC}{\affiliation{Oskar Klein Centre, Department of Astronomy, Stockholm University, Albanova University Centre, SE-106 91 Stockholm, Sweden}}
\newcommand{\OAPD}{\affiliation{INAF-Osservatorio Astronomico di Padova, Vicolo dell'Osservatorio 5, I-35122 Padova, Italy}}
\newcommand{\Caltech}{\affiliation{Cahill Center for Astronomy and Astrophysics, California Institute of Technology, Mail Code 249-17, Pasadena, CA 91125, USA}}
\newcommand{\GSFC}{\affiliation{Astrophysics Science Division, NASA Goddard Space Flight Center, Mail Code 661, Greenbelt, MD 20771, USA}}
\newcommand{\UMD}{\affiliation{Joint Space-Science Institute, University of Maryland, College Park, MD 20742, USA}}
\newcommand{\UCB}{\affiliation{Department of Astronomy, University of California, Berkeley, CA 94720-3411, USA}}
\newcommand{\TTU}{\affiliation{Department of Physics, Texas Tech University, Box 41051, Lubbock, TX 79409-1051, USA}}
\newcommand{\STScI}{\affiliation{Space Telescope Science Institute, 3700 San Martin Drive, Baltimore, MD 21218-2410, USA}}
\newcommand{\UT}{\affiliation{University of Texas at Austin, 1 University Station C1400, Austin, TX 78712-0259, USA}}
\newcommand{\IoA}{\affiliation{Institute of Astronomy, University of Cambridge, Madingley Road, Cambridge CB3 0HA, UK}}
\newcommand{\QUB}{\affiliation{Astrophysics Research Centre, School of Mathematics and Physics, Queen's University Belfast, Belfast BT7 1NN, UK}}
\newcommand{\IPAC}{\affiliation{Spitzer Science Center, California Institute of Technology, Pasadena, CA 91125, USA}}
\newcommand{\JPL}{\affiliation{Jet Propulsion Laboratory, California Institute of Technology, 4800 Oak Grove Dr, Pasadena, CA 91109, USA}}
\newcommand{\Southampton}{\affiliation{Department of Physics and Astronomy, University of Southampton, Southampton SO17 1BJ, UK}}
\newcommand{\LANL}{\affiliation{Space and Remote Sensing, MS B244, Los Alamos National Laboratory, Los Alamos, NM 87545, USA}}
\newcommand{\Tsinghua}{\affiliation{Physics Department and Tsinghua Center for Astrophysics, Tsinghua University, Beijing, 100084, People's Republic of China}}
\newcommand{\NAOC}{\affiliation{National Astronomical Observatory of China, Chinese Academy of Sciences, Beijing, 100012, People's Republic of China}}
\newcommand{\Itagaki}{\affiliation{Itagaki Astronomical Observatory, Yamagata 990-2492, Japan}}
\newcommand{\Einstein}{\altaffiliation{Einstein Fellow}}
\newcommand{\Hubble}{\altaffiliation{Hubble Fellow}}
\newcommand{\CfA}{\affiliation{Center for Astrophysics \textbar{} Harvard \& Smithsonian, 60 Garden Street, Cambridge, MA 02138-1516, USA}}
\newcommand{\UA}{\affiliation{Steward Observatory, University of Arizona, 933 North Cherry Avenue, Tucson, AZ 85721-0065, USA}}
\newcommand{\MPIA}{\affiliation{Max-Planck-Institut f\"ur Astrophysik, Karl-Schwarzschild-Stra\ss{}e 1, D-85748 Garching, Germany}}
\newcommand{\DSFP}{\altaffiliation{LSSTC Data Science Fellow}}
\newcommand{\HCO}{\affiliation{Harvard College Observatory, 60 Garden Street, Cambridge, MA 02138-1516, USA}}
\newcommand{\Carnegie}{\affiliation{Observatories of the Carnegie Institute for Science, 813 Santa Barbara Street, Pasadena, CA 91101-1232, USA}}
\newcommand{\TAU}{\affiliation{School of Physics and Astronomy, Tel Aviv University, Tel Aviv 69978, Israel}}
\newcommand{\Edinburgh}{\affiliation{Institute for Astronomy, University of Edinburgh, Royal Observatory, Blackford Hill EH9 3HJ, UK}}
\newcommand{\Birmingham}{\affiliation{Birmingham Institute for Gravitational Wave Astronomy and School of Physics and Astronomy, University of Birmingham, Birmingham B15 2TT, UK}}
\newcommand{\Bath}{\affiliation{Department of Physics, University of Bath, Claverton Down, Bath BA2 7AY, UK}}
\newcommand{\CTIO}{\affiliation{Cerro Tololo Inter-American Observatory, National Optical Astronomy Observatory, Casilla 603, La Serena, Chile}}
\newcommand{\Potsdam}{\affiliation{Institut f\"ur Physik und Astronomie, Universit\"at Potsdam, Haus 28, Karl-Liebknecht-Str. 24/25, D-14476 Potsdam-Golm, Germany}}
\newcommand{\INPE}{\affiliation{Instituto Nacional de Pesquisas Espaciais, Avenida dos Astronautas 1758, 12227-010, S\~ao Jos\'e dos Campos -- SP, Brazil}}
\newcommand{\UNC}{\affiliation{Department of Physics and Astronomy, University of North Carolina, 120 East Cameron Avenue, Chapel Hill, NC 27599, USA}}
\newcommand{\Ohio}{\affiliation{Astrophysical Institute, Department of Physics and Astronomy, 251B Clippinger Lab, Ohio University, Athens, OH 45701-2942, USA}}
\newcommand{\AAS}{\affiliation{American Astronomical Society, 1667 K~Street NW, Suite 800, Washington, DC 20006-1681, USA}}
\newcommand{\MMT}{\affiliation{MMT and Steward Observatories, University of Arizona, 933 North Cherry Avenue, Tucson, AZ 85721-0065, USA}}
\newcommand{\Geneva}{\affiliation{ISDC, Department of Astronomy, University of Geneva, Chemin d'\'Ecogia, 16 CH-1290 Versoix, Switzerland}}
\newcommand{\IUCAA}{\affiliation{Inter-University Center for Astronomy and Astrophysics, Post Bag 4, Ganeshkhind, Pune, Maharashtra 411007, India}}
\newcommand{\CMU}{\affiliation{Department of Physics, Carnegie Mellon University, 5000 Forbes Avenue, Pittsburgh, PA 15213-3815, USA}}
\newcommand{\NAOJ}{\affiliation{Division of Science, National Astronomical Observatory of Japan, 2-21-1 Osawa, Mitaka, Tokyo 181-8588, Japan}}
\newcommand{\IfA}{\affiliation{Institute for Astronomy, University of Hawai`i, 2680 Woodlawn Drive, Honolulu, HI 96822-1839, USA}}
\newcommand{\UCSC}{\affiliation{Department of Astronomy and Astrophysics, University of California, Santa Cruz, CA 95064-1077, USA}}
\newcommand{\Purdue}{\affiliation{Department of Physics and Astronomy, Purdue University, 525 Northwestern Avenue, West Lafayette, IN 47907-2036, USA}}
\newcommand{\Princeton}{\affiliation{Department of Astrophysical Sciences, Princeton University, 4 Ivy Lane, Princeton, NJ 08540-7219, USA}}
\newcommand{\Moore}{\affiliation{Gordon and Betty Moore Foundation, 1661 Page Mill Road, Palo Alto, CA 94304-1209, USA}}
\newcommand{\Durham}{\affiliation{Department of Physics, Durham University, South Road, Durham, DH1 3LE, UK}}
\newcommand{\JHU}{\affiliation{Department of Physics and Astronomy, The Johns Hopkins University, 3400 North Charles Street, Baltimore, MD 21218, USA}}
\newcommand{\Toronto}{\affiliation{David A.\ Dunlap Department of Astronomy and Astrophysics, University of Toronto,\\ 50 St.\ George Street, Toronto, Ontario, M5S 3H4 Canada}}
\newcommand{\Duke}{\affiliation{Department of Physics, Duke University, Campus Box 90305, Durham, NC 27708, USA}}
\newcommand{\NCU}{\affiliation{Graduate Institute of Astronomy, National Central University, 300 Jhongda Road, 32001 Jhongli, Taiwan}}
\newcommand{\Columbia}{\affiliation{Department of Physics and Columbia Astrophysics Laboratory, Columbia University, Pupin Hall, New York, NY 10027, USA}}
\newcommand{\Flatiron}{\affiliation{Center for Computational Astrophysics, Flatiron Institute, 162 5th Avenue, New York, NY 10010-5902, USA}}
\newcommand{\CIERA}{\affiliation{Center for Interdisciplinary Exploration and Research in Astrophysics and Department of Physics and Astronomy, \\Northwestern University, 1800 Sherman Avenue, 8th Floor, Evanston, IL 60201, USA}}
\newcommand{\GeminiNorth}{\affiliation{Gemini Observatory, 670 North A`ohoku Place, Hilo, HI 96720-2700, USA}}
\newcommand{\Keck}{\affiliation{W.~M.~Keck Observatory, 65-1120 M\=amalahoa Highway, Kamuela, HI 96743-8431, USA}}
\newcommand{\UW}{\affiliation{Department of Astronomy, University of Washington, 3910 15th Avenue NE, Seattle, WA 98195-0002, USA}}
\newcommand{\DiRAC}{\altaffiliation{DiRAC Fellow}}
\newcommand{\USask}{\affiliation{Department of Physics \& Engineering Physics, University of Saskatchewan, 116 Science Place, Saskatoon, SK S7N 5E2, Canada}}
\newcommand{\Thacher}{\affiliation{Thacher School, 5025 Thacher Road, Ojai, CA 93023-8304, USA}}
\newcommand{\Rutgers}{\affiliation{Department of Physics and Astronomy, Rutgers, the State University of New Jersey,\\136 Frelinghuysen Road, Piscataway, NJ 08854-8019, USA}}
\newcommand{\FSU}{\affiliation{Department of Physics, Florida State University, 77 Chieftan Way, Tallahassee, FL 32306-4350, USA}}
\newcommand{\Melbourne}{\affiliation{School of Physics, The University of Melbourne, Parkville, VIC 3010, Australia}}
\newcommand{\ASTROthreeD}{\affiliation{ARC Centre of Excellence for All Sky Astrophysics in 3 Dimensions (ASTRO 3D)}}
\newcommand{\Stromlo}{\affiliation{Mt.\ Stromlo Observatory, The Research School of Astronomy and Astrophysics, Australian National University, ACT 2601, Australia}}
\newcommand{\NCPAS}{\affiliation{National Centre for the Public Awareness of Science, Australian National University, Canberra, ACT 2611, Australia}}
\newcommand{\TAMU}{\affiliation{Department of Physics and Astronomy, Texas A\&M University, 4242 TAMU, College Station, TX 77843, USA}}
\newcommand{\Mitchell}{\affiliation{George P.\ and Cynthia Woods Mitchell Institute for Fundamental Physics \& Astronomy, College Station, TX 77843, USA}}
\newcommand{\ESO}{\affiliation{European Southern Observatory, Alonso de C\'ordova 3107, Casilla 19, Santiago, Chile}}
\newcommand{\ICE}{\affiliation{Institute of Space Sciences (ICE, CSIC), Campus UAB, Carrer
de Can Magrans, s/n, E-08193 Barcelona, Spain}}
\newcommand{\IEEC}{\affiliation{Institut d'Estudis Espacials de Catalunya, Gran Capit\`a, 2-4, Edifici Nexus, Desp.\ 201, E-08034 Barcelona, Spain}}
\newcommand{\Warwick}{\affiliation{Department of Physics, University of Warwick, Gibbet Hill Road, Coventry CV4 7AL, UK}}
\newcommand{\Macquarie}{\affiliation{School of Mathematical and Physical Sciences, Macquarie University, NSW 2109, Australia}}
\newcommand{\AAARC}{\affiliation{Astronomy, Astrophysics and Astrophotonics Research Centre, Macquarie University, Sydney, NSW 2109, Australia}}
\newcommand{\Capodimonte}{\affiliation{INAF -- Capodimonte Astronomical Observatory, Salita Moiariello 16, I-80131 Napoli, Italy}}
\newcommand{\INFNNapoli}{\affiliation{INFN -- Napoli, Strada Comunale Cinthia, I-80126 Napoli, Italy}}
\newcommand{\ICRANet}{\affiliation{ICRANet, Piazza della Repubblica 10, I-65122 Pescara, Italy}}
\newcommand{\MSU}{\affiliation{Center for Data Intensive and Time Domain Astronomy, Department of Physics and Astronomy,\\Michigan State University, East Lansing, MI 48824, USA}}
\newcommand{\SETI}{\affiliation{SETI Institute,
339 Bernardo Ave, Suite 200, Mountain View, CA 94043, USA}}
\newcommand{\IAIFI}{\affiliation{The NSF AI Institute for Artificial Intelligence and Fundamental Interactions}}
\newcommand{\Catalyst}{\altaffiliation{LSSTC Catalyst Fellow}}
\newcommand{\INAOE}{\affiliation{Instituto Nacional de Astrof{\'i}sica, {\'O}ptica y Electr{\'o}nica (INAOE CONACyT), Luis E. Erro 1, 72840, Tonantzintla, Puebla, Mexico}}

%% file: authors.tex
\author[0000-0002-7015-3446]{Nicol\'as Meza Retamal}
\UCD
\author[0000-0002-7937-6371]{Yize Dong \begin{CJK*}{UTF8}{gbsn}(董一泽)\end{CJK*}}
\UCD
\author[0000-0002-4924-444X]{K. Azalee Bostroem}
\Catalyst\UA
\author[0000-0001-8818-0795]{Stefano Valenti}
\UCD
\author[0000-0002-1296-6887]{Llu\'is Galbany}
\ICE \IEEC
\author[0000-0002-0744-0047]{Jeniveve Pearson}
\UA
\author[0000-0002-0832-2974]{Griffin Hosseinzadeh}
\UA
\author[0000-0003-0123-0062]{Jennifer E. Andrews}
\GeminiNorth
\author[0000-0003-4102-380X]{David J. Sand}
\UA
\author[0000-0001-5754-4007]{Jacob E. Jencson}
\UA
\author[0000-0003-0549-3281]{Daryl Janzen}
\USask
\author[0000-0001-9589-3793]{Michael~J. Lundquist}
\Keck
\author[0000-0003-2744-4755]{Emily T. Hoang}
\UCD
\author[0000-0003-2732-4956]{Samuel Wyatt}
\UW
\author[0000-0001-6272-5507]{Peter J.\ Brown}
\TAMU
\author[0000-0003-4253-656X]{D.\ Andrew Howell}
\LCO\UCSB
\author[0000-0001-9570-0584]{Megan Newsome}
\LCO\UCSB
\author[0000-0003-0209-9246]{Estefania Padilla Gonzalez}
\LCO\UCSB
\author[0000-0002-7472-1279]{Craig Pellegrino}
\LCO\UCSB
\author[0000-0003-0794-5982]{Giacomo Terreran}
\LCO\UCSB
\author[0000-0003-3642-5484]{Vladimir Kouprianov}
\UNC
\author[0000-0002-1125-9187]{Daichi Hiramatsu}
\CfA\IAIFI
\author[0000-0001-8738-6011]{Saurabh W.\ Jha}
\Rutgers
\author[0000-0001-5510-2424]{Nathan Smith}
\UA
\author[0000-0002-6703-805X]{Joshua Haislip}
\UNC
\author[0000-0002-5060-3673]{Daniel E.\ Reichart}
\UNC
\author[0000-0002-4022-1874]{Manisha Shrestha}
\UA
\author[0000-0002-3642-9146]{F. Fabi\'an Rosales-Ortega}
\INAOE
\author[0000-0001-5955-2502]{Thomas G. Brink}
\UCB
\author[0000-0003-3460-0103]{Alexei V. Filippenko}
\UCB
\author[0000-0002-2636-6508]{WeiKang Zheng}
\UCB
\author[0000-0002-6535-8500]{Yi Yang}
\UCB
\Tsinghua